\documentclass[paper]{JHEP}
\usepackage{epsfig}
\usepackage{amsmath}
\usepackage{amssymb}
\usepackage{multirow}
\def\beq{\begin{equation}}
\def\beqn{\begin{eqnarray}}
\def\eeq{\end{equation}}
\def\eeqn{\end{eqnarray}}

\def \ap {{\cal A}_+}
\def \am {{\cal A}_-}
\def \real {\,\mathrm{Re}}
\newcommand\sss{\scriptscriptstyle\rm}

\newcommand\as{\alpha_{\sss S}}

\newcommand\mydot{\!\cdot\!}

\newcommand\mat{{\cal M}}

\newcommand{\mh}{{\ensuremath m_{H^-}}}
\newcommand{\gev}{{\ensuremath\rm GeV}}

\preprint{LPSC-09-188, NIKHEF-2009-013\\
ITFA-2009-21, FERMILAB-PUB-09-626-T\\
 IPPP/09/95, DCPT/09/190\\
CERN-TH/2009-251, ITF-UU-09/32\\
}
\title{Charged Higgs boson production in association with a top quark in MC@NLO}
\author{C. Weydert$^a$, S. Frixione$^{b,c}$, M. Herquet$^d$, M. Klasen$^a$\\
  $^a$Laboratoire de Physique Subatomique et de Cosmologie, UJF, CNRS/IN2P3, INPG, 53 avenue des Martyrs, 38026 Grenoble cedex, France\\
  $^b$PH Department, TH unit, CERN, CH-1211 Geneva 23, Switzerland\\
  $^c$ITPP, EPFL, CH-1015 Lausanne, Switzerland\\
  $^d$Nikhef Theory Group,  Science Park 105, 1098 XG Amsterdam, The Netherlands\\
  E-mail: \email{weydert@lpsc.in2p3.fr}, \email{Stefano.Frixione@cern.ch}, \email{mherquet@nikhef.nl}, \email{klasen@lpsc.in2p3.fr}}
\author{E. Laenen$^{d,e,f}$, T. Plehn$^g$, G. Stavenga$^{f,h}$, C. D. White$^i$\\
  $^e$Institute for Theoretical Physics, University of Amsterdam,Valckenierstraat 65, 1018 XE Amsterdam, The Netherlands\\
  $^f$Institute for Theoretical Physics, Utrecht University, Leuvenlaan 4, 3584 CE Utrecht, The Netherlands\\
  $^g$Institut f\"{u}r Theoretische Physik, Universit\"{a}t Heidelberg, Germany\\
  $^h$Fermi National Accelerator Laboratory, MS106, P.O. Box 500, IL 60510, U.S.A.\\
  $^i$Institute for Particle Physics Phenomenology, Department of Physics, Durham University, Durham DH1 3LE, United Kingdom\\
  E-mail: \email{Eric.Laenen@nikhef.nl}, \email{Plehn@uni-heidelberg.de}, \email{stavenga@fnal.gov}, \email{c.d.white@durham.ac.uk} 
}

\abstract{We discuss the calculation of charged Higgs boson production in association with a top quark in the MC@NLO framework for combining NLO matrix elements with a parton shower. The process is defined in a model independent manner for wide applicability, and can be used if
the charged Higgs boson mass is either greater or less than the mass of the top quark. For the latter mass region, care is needed
 in defining the charged Higgs production mode due to interference with top pair production. We give a suitable definition applicable in an NLO (plus parton shower) context, and present example results for the LHC.}
\enlargethispage*{1000pt}

\begin{document}
\section{Introduction}
\label{sec:introduction}

Even in the absence of any definitive experimental evidence, the Higgs mechanism of spontaneous symmetry breaking \cite{Higgs:1964pj,Higgs:1966ev,Englert:1964et,Guralnik:1964eu,Kibble:1967sv} remains a very promising candidate to explain the existence of massive weak gauge bosons. This mechanism is implemented in its minimal version within the Standard Model (SM), i.e. using a single $SU(2)_L$ Higgs doublet. Yet, more complex scenarios involving additional scalar fields remain possible and could display interesting properties such as, for example, new sources of $CP$ violation, and/or embedding in more complex models like the Minimal Supersymmetric Standard Model (MSSM). In this context, existing and future high energy colliders will have to determine not only if the elusive Higgs particle exists, but also if the observed scalar sector is minimal or not. Any conclusive answer to the last question strongly relies upon the possibility of observing a charged Higgs boson. Indeed, the discovery of such a particle would clearly imply the presence of additional non trivial weak multiplet(s) in the scalar potential responsible for spontaneous symmetry breaking.

If the charged Higgs boson is lighter than the top quark, the most promising production mode at hadron colliders (see e.g.~\cite{Djouadi:2005gj} 
for a recent review) is through the top quark decay $t\rightarrow H^+ b$ (or $\bar{t}\rightarrow H^- \bar{b}$). If it is larger, as suggested by indirect measurements such as the $b\rightarrow s\gamma$ branching ratio~\cite{Misiak:2006zs}, the most promising process is the direct production in association with a single top quark, which is the focus of the present work. Other production modes for the charged Higgs boson, such as $s$-channel single, pair or associated production processes, are less favorable for discovery in most models. 

In order to define as precisely as possible an isolation strategy and (if a charged Higgs boson happens to be actually observed) in order to render possible the extraction of physical parameters such as its coupling to heavy fermions, accurate predictions are necessary at the fully 
exclusive level for this channel. For similar Standard Model processes involving a single top quark (which can play the role of backgrounds to the considered channel), the current state of the art is the combination of NLO parton-level matrix elements with Monte Carlo event generators. Those generators use parton shower algorithms to simulate the effect of further soft and collinearly enhanced radiation, as well as modelling hadronization effects and the underlying event. One such approach for combining NLO matrix elements with parton showers is the MC@NLO algorithm of~\cite{Frixione:2002ik}. All of the three single top production modes have all already been implemented in this framework~\cite{Frixione:2005vw,Frixione:2008yi}, including angular correlations using the method described in~\cite{Frixione:2007zp} (for a recent study of angular correlations in top quark production, see~\cite{Motylinski:2009kt}).  It is thus natural to implement the production of a charged Higgs boson ($H^\pm$) with an accompanying single top quark, first calculated at NLO in~\cite{Zhu:2001nt,Plehn:2002vy}. Single top production in both the $s$- and $t$-channel modes was also recently implemented in the POWHEG framework for combining NLO matrix elements with parton showers~\cite{Alioli:2009je}.

As we will see, $H^\pm t$ production is theoretically similar to the $Wt$ mode (essentially, one replaces the $W$ boson in all $Wt$ Feynman diagrams with a charged Higgs boson). This creates an extra motivation for studying $H^\pm$ production in a framework which combines NLO matrix elements with parton showers. Indeed, it is well known that $Wt$ production  at NLO mixes with top pair production at LO, followed by decay of one of the final state top particles. Thus, the meaning of the $Wt$ mode itself, and by analogy the $H^\pm$ mode when $m_{H^\pm}<m_t$, becomes a matter of careful definition, which must be applicable in an experimental context. This then demands calculations which are at least as complex as MC@NLO, in that the problem occurs only at NLO and beyond, and only in a fully exclusive, all-orders computation can a suitable definition of the considered mode be tested within an experimental context.

The structure of the paper is as follows. In section \ref{NLO} we outline the computation of the charged Higgs production mode at NLO in QCD, 
together with its implementation in the MC@NLO framework. In section \ref{interference}, we describe how to define the $H^-t$ mode for $m_{H^-}<m_t$ in two different ways (whose comparison measures interference contributions), leaning on the preceding discussion of $Wt$ production in~\cite{Frixione:2008yi}. In section \ref{results} we present example results, also discussing the transition region where $m_{H^-}\simeq m_t$. We conclude in section \ref{conclusion}.

\section{$H^- t$ production at NLO}
\label{NLO}
In this section we describe our calculation of the $H^-t$ process up to NLO in QCD perturbation theory. In order to check our results, we performed the calculation using two different methods for dealing with infrared divergences in the real and virtual contributions. Firstly, we used the Catani-Seymour algorithm of~\cite{Catani:1996vz,Catani:2002hc}. Given that this calculation has not been performed before, we present some salient details in what follows. Secondly, we used the FKS subtraction formalism of~\cite{Frixione:1995ms,Frixione:1997np}. This is the formalism required for the implementation of NLO results within MC@NLO, and as such has already been used in previous processes. Hence, the details are extremely similar to the case of $Wt$ production considered in~\cite{Frixione:2008yi}, and we do not present many of them here. We start by discussing the LO result, which will also be useful in introducing notation, in the following section.
\subsection{Born computation}
\begin{figure}
\begin{center}
\scalebox{0.8}{\includegraphics{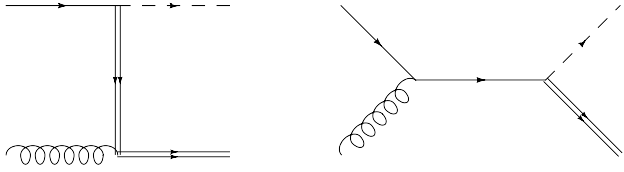}}
\caption{Leading order diagrams for single top production (double line) in association with a charged Higgs boson (dashed line).}
\label{Bornfig}
\end{center}
\end{figure}

The LO Feynman diagrams for charged Higgs production are shown in figure \ref{Bornfig}. From now on, we consider explicitly $H^-$ production (the case of $H^+$ production being the same, as shown for the $Wt$ case in~\cite{Frixione:2008yi}). Note that, as in~\cite{Frixione:2008yi}, we assume a five flavor scheme~\cite{Aivazis:1993pi} for the quark sector, where the $b$ quark is massless. The case where $b$ quarks are generated explicitly by initial state gluon splitting was considered in~\cite{Plehn:2002vy}.
We label momenta as follows:
\begin{equation}
b(p_1)+g(p_2)\rightarrow t(k_1)+H^-(k_2).
\label{Borneq}
\end{equation}
One must define the coupling of the charged Higgs boson to fermions, which is clearly a matter of convention. It is of most practical use to 
leave this coupling model-independent, and we choose to write the Higgs-fermion vertex as follows:
\begin{equation}
G_{H^-U\bar{D}}=iV_{U\bar{D}}[A_{U\bar{D}}-B_{U\bar{D}}\gamma_5],
\label{coupling}
\end{equation}
where $U$ and $D$ are up- and down- type quarks respectively, and we have explicitly factored out the corresponding CKM matrix element 
$V_{U\bar{D}}$. To simplify notation from now on, we consider the case where the $H^-$ couples only to a $t$ and a $b$ quark 
(by far the most dominant process), and write 
\begin{equation}
G_{H^-tb}=iV_{tb}[a-b\gamma_5],
\label{coupling2}
\end{equation}
where $a$ and $b$ may be complex in general. When presenting results later in the paper, we will use the specific example of a type-II two-Higgs doublet model, in which $a$ and $b$ are given as in eq.~(\ref{aandb}) of appendix~\ref{renormy}, and depend explicitly on the top and bottom quark masses. In the matrix element, all quarks other than the top are treated as massless, including the $b$ quark. In both the Born and NLO calculations, the pure scalar and pure pseudoscalar couplings add incoherently i.e. the total cross-section is proportional
to $|a|^2+|b|^2$. A comment is in order regarding the use of the massless $b$ quark approximation. The reader may worry
about the fact that $m_b$ dependence is kept in the Yukawa couplings, but neglected in the matrix elements. In actual fact
there is no inconsistency, as discussed clearly in~\cite{Plehn:2002vy,Berger:2003sm}. Keeping the $b$ mass in the Yukawa coupling merely corresponds to keeping only the leading $m_b$ behavior.
\subsection{NLO computation}
As stated above, we performed the calculation at NLO accuracy using two different subtraction formalisms for dealing with infrared divergences in order to check our results. Before discussing these in more detail, some remarks are in order regarding the treatment of ultraviolet divergences in the couplings and heavy quark mass.
\subsubsection{Coupling and mass renormalization}
We evaluated all one-loop diagrams using dimensional regularization in $d=4-2\epsilon$ dimensions. Up to NLO, one finds both double and single poles in $\epsilon$, arising from infra-red (IR) and ultraviolet (UV) divergences. The former cancel in the sum of virtual and real corrections (or, in the case of initial state collinear divergences, are removed by counterterms), as described in the next section. UV divergences are removed by renormalization of the strong and Yukawa couplings, and of the top quark mass.

As in~\cite{Frixione:2008yi}, we modify the $\overline{\text{MS}}$-scheme QCD coupling such that the top quark loop contribution is subtracted on-shell. In this scheme~\cite{Collins:1978wz}, the top quark virtual contributions decouple in the limit of small external momentum. Specifically, one has
\begin{equation}
g_s\rightarrow g_s(\mu_R^2)\left[1+\left(\frac{\alpha_S(\mu_R)}{8\pi}\right)\left(-\frac{1}{\epsilon}+\gamma_E-\ln4\pi\right)\beta_0\left(\frac{\mu^2}{\mu_R^2}\right)^\epsilon+\left(\frac{\alpha_S(\mu_R^2)}{8\pi}\right)\frac{2}{3}\ln\left(\frac{\mu_R^2}{m_t^2}\right)\right],
\label{gsrenorm}
\end{equation}
where $\mu_F$ and $\mu_R$ are the factorization and renormalization scales respectively. Furthermore, $\beta_0=(11C_A-2n_f)/3$, with $n_f$ the number of light flavors (here five) plus one. The QCD coupling then satisfies:
\begin{equation}
\mu_R^2\frac{dg_s(\mu_R^2)}{d\mu_R^2}=-g_s(\mu_R^2)\left(\frac{\alpha_S(\mu_R^2)}{8\pi}\right)\left(\beta_0+\frac{2}{3}\right)+{\cal O}(g_s^5),
\label{RGE}
\end{equation}
so that the top quark loop contribution is indeed removed.

The renormalization of the top quark mass in the on-shell scheme is given by
\begin{align}
m_t\rightarrow m_t&+\delta m_t=\notag\\
&m_t\left[1+\left(\frac{\alpha_S(\mu_R^2)}{4\pi}\right)C_F\left(-\frac{3}{\epsilon}+3\gamma_E-3\ln4\pi-4-3\ln\frac{\mu_R^2}{m_t^2}\right)\right].
\label{mrenorm}
\end{align}
The renormalization of the Yukawa coupling $y$ (which represents either $a$ or $b$ in eq.~(\ref{coupling2})) to a quark of mass $m$ is related to the appropriate mass counterterm via (see appendix~\ref{renormy})
\begin{equation}
\delta y=\frac{\delta m}{m},
\label{yrenorm}
\end{equation}
where the renormalized quark mass appears in the denominator on the right-hand side. Note that this result is independent of the mass (since $\delta m\propto m$), and also that the mass counterterms appearing in the renormalization of both the heavy quark mass and Yukawa couplings must be in the same renormalization scheme (in our case the on-shell scheme). 
\subsubsection{Catani-Seymour subtraction}
Here we briefly summarize the NLO calculation of the $H^-t$ process in the Catani-Seymour dipole formalism of~\cite{Catani:1996vz,Catani:2002hc}. Our terminology and notation is similar to those papers, to which we refer the reader for more details.

In the dipole formalism, as in any other subtraction formalism, the
 real and virtual contributions are dealt with so as to render them 
 separately finite and thus numerically integrable. This is achieved
 through a subtraction of the leading phase-space divergences 
 in the former, and of the infrared $1/\epsilon^k$ poles in the
 latter; we shall give more details in the following. The total NLO partonic cross-section has the form
\begin{align}
\sigma^{\text{NLO}}(p_1,p_2;\mu_F^2)&=\sigma^{\text{NLO}(2)}(p_1,p_2)+\sigma^{\text{NLO}(3)}(p_1,p_2)+\int_0^1 dx_1\sigma^{\text{NLO}}(x_1;x_1p_1,p_2;\mu_F^2)\notag\\
&\quad+\int_0^1 dx_2\sigma^{\text{NLO}}(x_2;p_1,x_2p_2;\mu_F^2),
\label{CStot}
\end{align}
where we have explicitly denoted the dependence on the initial momenta $p_i$, and $\mu_F$ is the factorization scale. The first term on the right-hand side of eq.~(\ref{CStot}) has $2\rightarrow 2$ kinematics and is comprised of the virtual corrections and subtraction term. The second term has $2\rightarrow 3$ kinematics, and includes the regularized real corrections. The final two terms constitute a finite remainder left after factorization of initial state collinear singularities, and involve integrals over the longitudinal momentum fractions $x_i$ of the initial state partons.

In more detail, the regularized virtual corrections have the form
\begin{align}
\sigma^{\text{NLO}(2)}(p_1,p_2)&=\int d\Phi^{(2)}(p_1,p_2)\Big[\frac{1}{{\cal N}_g{\cal N}_b}2\,\text{Re}\,\left[{\cal M}_{\text{1-loop}}\,{\cal M}^\dag_{\text{Born}}\right]\notag\\
&\quad+\langle t,H;g,b|{\bf I}(\epsilon)|t,H;g,b\rangle\Big]_{\epsilon=0},
\label{virt}
\end{align}
where ${\cal N}_a$ is the number of color degrees of freedom associated with incoming parton $a$, and $d\Phi^{(2)}$ the phase-space of the two final state particles (including spin averaging). Furthermore, the second term denotes the subtraction operator ${\bf I}$, which is matrix-valued in color space and acts on the color vectors $|t,H;g,b\rangle$ associated with the particles entering the Born interaction. The subtraction operator may be further decomposed as
\begin{equation}
{\bf I}(\epsilon)={\bf I}_2(\epsilon,\mu^2;m_t,k_2)+{\bf I}_{b}(\epsilon,\mu^2;k_2,m_t,p_1)+{\bf I}_{g}(\epsilon,\mu^2;k_2,m_t,p_2)+{\bf I}_{bg}(\epsilon,\mu^2;p_1,p_2),
\label{Iterm}
\end{equation}
where ${\bf I}_{a}$ includes the contribution from a gluon exchanged (across the final state cut in the squared amplitude) between parton $a$ and the top quark, ${\bf I}_{bg}$ that from a gluon exchanged between the initial state partons, and ${\bf I}_2$ from gluons exchanged between the final state particles (this is zero for the process considered here, given that the Higgs boson is a color singlet). 

The real emission term in eq.~(\ref{CStot}) has the schematic form
\begin{equation}
\sigma^{\text{NLO}(3)}(p_1,p_2)=\int d\Phi^{(3)}\left[\frac{1}{{\cal N}_g{\cal N}_b}\left|{\cal M}_3(k_1,k_2,k_3;p_1,p_2)\right|^2-\sum_{\text{dipoles}}{\cal D}(k_1,k_2,k_3;p_1,p_2)\right],
\label{real}
\end{equation}
where $k_3$ is the momentum of the extra parton at NLO, $d\Phi^{(3)}$ the phase-space of the final state particles, and ${\cal M}_3$ the real emission amplitude consisting of the following four subprocesses:

(a): $b(p_1)+g(p_2)\rightarrow t(k_1)+H^-(k_2)+g(k_3)$;

(b): $g(p_1)+g(p_2)\rightarrow t(k_1)+H^-(k_2)+\bar{b}(k_3)$;

(c): $\bar{q}/q(p_1)+b(p_2)\rightarrow t(k_1)+H^-(k_2)+\bar{q}/q(k_3)$;

(d): $\bar{q}(p_1)+q(p_2)\rightarrow t(k_1)+H^-(k_2)+\bar{b}(k_3)$.

Here $q$ denotes a generic quark species (which may be a $b$ quark). Note, however, that process (a) can be obtained from (b) by crossing. Also, processes (c) and (d) interfere when $q=b$, and care must be taken such that this is correctly dealt with. The contribution from these processes is in any case negligible in practice, due to the smallness of the $b$ quark parton density. When $q\neq b$, process (d) is finite as $\epsilon\rightarrow 0$.

The second term on the right-hand side of eq.~(\ref{real}) contains a sum over Catani-Seymour dipoles, representing all possible gluon exchanges across the final state cut, and collinear splittings of initial state particles. Using the standard nomenclature of~\cite{Catani:1996vz,Catani:2002hc} with upper and lower indices denoting initial and final state particles respectively, one may write these as ${\cal D}^{\alpha\beta,\gamma}$ and ${\cal D}^{\alpha\beta}_\gamma$ for an emitter $\alpha$ emitting particle $\beta$, and $\gamma$ the spectator particle. There are also dipoles ${\cal D}^{\gamma}_{\beta\alpha}$ corresponding to final state emitters and initial state spectators. For the present process there are no dipoles involving final state emitters and spectators, due to the final state color singlet Higgs particle. A complete classification of dipoles is then
\begin{itemize}
\item Process (a): $\sum_{\text{dipoles}}={\cal D}^{gg,b}+{\cal D}^{bg,g}+{\cal D}^{gg}_t+{\cal D}^{bg}_t+{\cal D}^g_{gt}+{\cal D}^b_{gt}$;
\item Process (b): $\sum_{\text{dipoles}}={\cal D}^{g_1b,g_2}+{\cal D}^{g_2b,g_1}+{\cal D}^{g_1b}_t+{\cal D}^{g_2b}_{t}$;
\item Process (c): $\sum_{\text{dipoles}}={\cal D}^{qq,b}+{\cal D}^{qq}_t$,
\end{itemize}
where $g_{1,2}$ denote the two initial state gluons in process (b).

Finally, the last term of eq.~(\ref{CStot}) (i.e. the finite remainder after collinear factorization) has the general expression
\begin{align}
\int_0^1dx_1\sigma^{\text{NLO}(2)}&(x_1;x_1p_1,p_2;\mu_F^2)=\sum_{i'}\int_0^1dx_1\int d\Phi^{(2)}(x_1p_1,p_2)\notag\\
&\quad\langle k_1,k_2;x_1p_1,p_2\left|\left({\bf K}^{i,i'}(x_1)+{\bf P}^{i,i'}(x_1p_1,x,\mu_F^2)\right)\right|k_1,k_2;x_1p_1,p_2\rangle.
\label{colterm}
\end{align}
for the initial state parton whose momentum is $p_1$, with a similar expression for $p_2$. In this formula, $i$ labels the parton emerging from the hadron, and $i'$ a particle from the collinear splitting which interacts with the second incoming particle. The functions ${\bf K}^{i,i'}$ and ${\bf P}^{i,i'}$ (matrix-valued in color space and evaluated between color vectors) are process-independent, and involve products of ingoing and outgoing momenta.
\subsubsection{FKS subtraction}
The FKS calculation is extremely similar to that carried out for $Wt$ production in~\cite{Frixione:2008yi}, and thus we do not repeat the discussion here (the calculation is also similar for both $m_{H^-}>m_t$ and $m_{H^-}<m_t$). We used FORM~\cite{Vermaseren:2000nd} for both the virtual and real corrections.

The virtual corrections in our two calculations were compared analytically, and found to agree exactly. The real corrections were compared numerically, and agreement found within numerical uncertainties. As a further check, we have compared the numerical implementation of our total NLO results with that of~\cite{Plehn:2002vy}, which is available in the publicly available Prospino2.0 package~\cite{Beenakker:1996ed,Prospino2}. Such checks are possible only for $m_{H^-}>m_t$, where one does not have to worry about interference with top pair production. For the total NLO cross-section, we find agreement within a percent for a range of charged Higgs boson masses between 300 GeV and 1 TeV. Note that in order to compare with the calculation of~\cite{Plehn:2002vy}, we changed the renormalization of the Yukawa coupling to the $\overline{\text{MS}}$ scheme rather than the on-shell scheme, as detailed in that paper. Checks were also performed for $\mu_F\neq\mu_R$ (to explicitly test terms involving logarithms of the form $\log(\mu_F/\mu_R$)), and a similar level of agreement found.  

Having checked the implementation of our NLO calculation, the FKS code was then interfaced with MC@NLO as prescribed in~\cite{Frixione:2002ik}. Note that there is a technical subtlety regarding the collinear limits for partonic subprocesses involving gluon branchings. In the case of the FKS formalism, one introduces additional universal splitting kernels (as described in~\cite{Frixione:1995ms} and applied also in~\cite{Frixione:2008yi}). See appendix~\ref{app-hel} for details.
\subsection{Implementation in MC@NLO}
\label{MC@NLO}
In this section, we describe how the calculation of the previous section can be implemented in the MC@NLO framework for combining NLO matrix elements with a parton shower algorithm. Given that several processes have already been implemented in this formalism, we refer the reader to~\cite{Frixione:2002ik,Frixione:2005vw,Frixione:2003ei} for more technical details, and here briefly describe only those aspects that are relevant for $H^-t$ production. Furthermore, the implementation of the present process is, as remarked in the previous sections, extremely similar to that of $Wt$ production considered in~\cite{Frixione:2008yi}. 

The MC@NLO algorithm first presented in~\cite{Frixione:2002ik} gives a systematic procedure for combining a NLO matrix element with a parton shower, in such a way that double counting of radiation is avoided. This is achieved through the definition of the so-called Monte Carlo
 subtraction terms, that are designed to cancel exactly the contributions
 at next-to-leading order to the cross section of interest, given by the
 parton shower. The MC subtraction terms are factorized into 
 process-independent kernels, that depend solely on how the Monte Carlo
 treats the collinear and soft emissions (e.g. through the definition 
 of the shower variables), times process-dependent short-distance
 cross sections. The latter essentially coincide with Born matrix elements,
 and are therefore available as part of the NLO parton-level cross
 sections. As stressed above, the kernels are MC-specific; for a given
 MC, only a handful of cases have to be considered, corresponding to
 initial- or final-state branchings, from a massless or a massive
 particle. All relevant computations have been carried out for
 the case where the MC is Fortran HERWIG (see in particular
 refs.~\cite{Frixione:2002ik,Frixione:2003ei,Frixione:2005vw}). This is now also the case for 
 Herwig++~\cite{Frixione}\footnote{Some processes interfaced
 with Herwig++ have already appeared in public MC@NLO codes~\cite{LatundeDada:2009rr,Papaefstathiou:2009sr,LatundeDada:2007jg}.};
 kernels relevant to initial-state radiation have also been computed
 for Pythia~\cite{Torrielli}. The subtraction terms have the schematic form
\begin{equation}
d\sigma\Big|_{\text{MC}}=\sum_i\sum_L\sum_l d\sigma_i^{(L,l)}\Big|_{\text{MC}},
\label{subterms}
\end{equation}
where $i$ labels different partonic subprocesses, $L$ the leg from which the extra parton is emitted (i.e. the parton appearing at NLO which is double counted by the parton shower), and $l$ denotes a given color structure. Furthermore, each partonic branching has a shower energy scale
\begin{equation}
E_0^2=|q_\alpha\cdot q_\beta|,
\label{sscale}
\end{equation}
where $q_{\alpha}$ and $q_\beta$ denote the 4-momenta of the color partners, one of which undergoes the branching. Thus, the shower scales are completely determined by the partonic subprocesses involved in the hard scattering.

Crucially, the partonic subprocesses and color structures involved in the $H^-t$ production process are the same as in $Wt$ production. Thus, the subtraction terms and shower scales for the present case do not have to be recalculated, and can be read off from section 3.3 of~\cite{Frixione:2008yi} (with the $W$ boson replaced by a charged Higgs). We refer the reader to that paper for details.

A given partonic subprocess may have more than one color flow, and it is then necessary to select one to feed to the parton shower Monte Carlo. We do this analogously to what was used in~\cite{Frixione:2008yi}, which can be summarized as follows. Where more than one color flow exists, we select one on a statistical basis, with each color flow weighted by its large $N_c$ approximation.

Note that in the present implementation, spin correlations have not yet been included in the decay of the top quark. These 
will be included at a later stage, using the technique of~\cite{Frixione:2007zp}.
\section{Interference with $t\bar{t}$ production}
\label{interference}
In the previous subsections, we have discussed the calculation of the $H^-t$ production process up to next-to-leading order in QCD, and its subsequent implementation in the MC@NLO framework for interfacing with a parton shower. In this discussion we ignored the fact that for $m_{H^-}<m_t$ a theoretical problem arises in that, at NLO and beyond, the single top process interferes with the production of a top quark pair, with decay of the antitop particle into a charged Higgs boson and $\bar{b}$ quark. We consider this issue, including how to recover a well-defined meaning of the $H^-t$ process, in this section. Our discussion leans heavily on the results obtained for the $Wt$ process in~\cite{Frixione:2008yi,White:2009yt}. 

\begin{figure}
\begin{center}
\scalebox{0.5}{\includegraphics{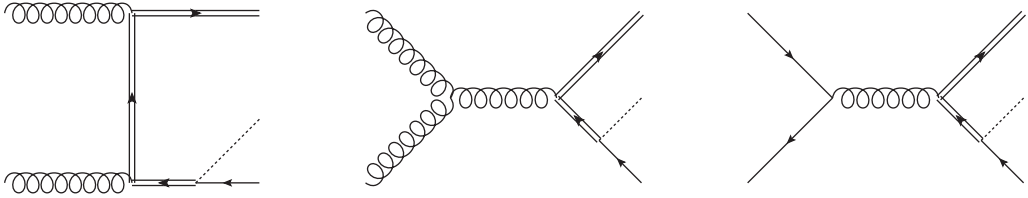}}
\caption{Subset of NLO real emission contributions to $H^-t$ production, consisting of top pair production with decay of the antitop quark to produce a charged Higgs boson and $\bar{b}$ quark.}
\label{idiags}
\end{center}
\end{figure}

The interference problem can be appreciated by considering figure~\ref{idiags}, which shows a subset of the NLO real-emission corrections to $H^-t$ production. The contribution from these diagrams grows as the invariant mass of the final state $H^-\bar{b}$ tends towards the top mass $m_t$, due to the propagator for the intermediate $\bar{t}$ quark which is moving on-shell. In practice this means that the NLO correction to the LO $H^-t$ inclusive cross-section is huge, with most of the correction due to the diagrams shown in figure~\ref{idiags}; this spoils the perturbation expansion. Note that this is not a problem when $m_{H^-}>m_t$, as in that case the kinematics of the final state $H^-\bar{b}$ system forbids the intermediate $\bar{t}$ from going on-shell.

However, as is clear from the figure and the fact that the problem occurs when the $\bar{t}$ is on-shell, the diagrams can also be thought of as the production of a top quark pair at leading order in QCD, with decay of the $\bar{t}$ particle producing a charged Higgs boson. That is, top pair and single top production interfere beyond LO in the single top cross-section, and the question arises of whether it is possible to separate the two processes, and thus maintain a meaningful definition of $H^-t$ production at higher orders in perturbation theory. It must be stressed that such a separation is an approximation for practical purposes only (i.e. $H^-t$ and $t\bar{t}$ production really do interfere at the quantum level), and is dependent on the experimental analysis cuts which are applied. However, and as we will see, such a separation is indeed possible subject to cuts, allowing the single and pair production processes to be added incoherently to a sufficiently good approximation.

There are a number of advantages that result from separating $H^-t$ from top pair production, in trying to accurately represent their sum. A theoretically rigorous superposition of single and top pair production would require at ${\cal O}(\alpha_S^2)$ that LO top pair production (with decay of the $\bar{t}$) be added to NLO $H^-t$ at the amplitude level. Such a procedure includes all interference effects, but neglects NLO contributions to the top pair process. The latter are known only in the narrow width approximation in which the $\bar{t}$ quark is produced on-shell, and are known to be large (of order $50\%$ for cuts used to isolate the top pair production cross-section). If one instead combines the single and top pair processes incoherently, the NLO corrections are included at the expense of the interference term. Importantly, for signal cuts used to isolate the single top process, the NLO corrections to top pair production are, in a well-defined sense, larger than the interference between the single and pair production processes. The most accurate description then results from adding the $H^-t$ and top pair matrix elements incoherently, thereby neglecting the interference term.

Such a discussion is not meaningful unless one has a means of quantifying the size of the interference between single top and top pair production. If this is indeed the case, then the systematic uncertainty due to interference effects can be compared with other uncertainties in a given analysis (such as scale variation), in order to ascertain whether or not it is legitimate to regard $H^-t$ and $t\bar{t}\rightarrow tH^-\bar{b}$ as incoherent processes.

The interference problem described above is by no means restricted to $H^-t$ production, but also occurs in other contexts such as single top quark production in association with a $W$ boson. Solutions to the problem in that context have been widely studied~\cite{Belyaev:1998dn,Tait:1999cf,Campbell:2005bb,Zhu:2002uj}. In~\cite{Frixione:2008yi} two definitions of the $Wt$ process were given which could be applied in a fully exclusive context i.e. within a parton shower framework. These definitions were called Diagram Removal (DR) and Diagram Subtraction (DS), and were subsequently implemented in MC@NLO. The definitions were designed so that the difference between them measured the degree of interference between $Wt$ and $t\bar{t}$ production, where the $\bar{t}$ in the latter case decays to a $W\bar{b}$ pair. Thus, the difference between the DR and DS results can be used to estimate the systematic uncertainty due to interference effects. This will not be small in general, and depends on the analysis cuts applied to the final state. 

A phenomenological analysis was carried out using these results in~\cite{White:2009yt}, which showed that for signal cuts used to isolate the $Wt$ signal, the interference with top pair production is indeed small with respect to other uncertainties in the analysis (most notably, the scale variation of the cross-section which is representative of the size of higher order corrections). This was also the case when $Wt$ and top pair production were themselves backgrounds to a third process (the specific example of Standard Model Higgs production followed by decay to a $W$ boson pair was considered). In such cases, the most accurate description of the relevant single and top pair production processes results from an incoherent sum of the two, for which one can use MC@NLO as detailed in those papers. The strong theoretical similarities between the $Wt$ and $H^-t$ amplitudes imply that the same definitions can be applied in the latter case, and that they will be similarly successful from the phenomenological point of view. This is indeed the case, as we will see in the following subsections. First, we recap the definitions of DR and DS, with the discussion tailored to the present context.
\subsection{Diagram Removal (DR) and Diagram Subtraction (DS)}
In DR, the diagrams of figure~\ref{idiags} (which we may call {\it doubly resonant} due to the intermediate top quark pair) are removed at the amplitude level. Thus, the squared NLO amplitude for $H^-t$ production has no interference with top pair production by construction. A potential difficulty of this approach is that both electroweak and QCD gauge invariance are violated. However, this was seen not to be a problem in practice (see~\cite{Frixione:2008yi} for a detailed discussion).

In DS, the cross-section for $H^-t$ production is modified by a local subtraction term, which effectively removes the contribution of doubly resonant diagrams point-by-point in phase space. This is a gauge-invariant procedure, although there is a degree of arbitrariness in the explicit construction of the subtraction term. Schematically, one has
\begin{equation}
d\sigma^{DS}_{H^-t}=d\sigma_{H^-t}-d\sigma^{sub}_{H^-t},
\label{DSdef}
\end{equation}
where the first term on the right-hand-side denotes the fully exclusive cross-section for $H^-t$ production (including doubly resonant diagrams), and the second the subtraction term. The latter must satisfy the following requirements:
\begin{enumerate}
\item When the invariant mass of the final state $H^-\bar{b}$ system satisfies $m_{H^-\bar{b}}=m_t$, the subtraction term must be equal to the fully exclusive cross-section for top pair production, with the $\bar{t}$  decaying to $H^-\bar{b}$.
\item The subtraction term must fall away sharply as $m_{H^-\bar{b}}$ moves away from the top mass.
\end{enumerate}
The first condition amounts to subtracting the doubly resonant contribution as required, and the second ensures that the diagrams in figure~\ref{idiags} contribute unmodified when the $\bar{t}$ is off-shell. These are the only two conditions that the subtraction term must satisfy, and following~\cite{Frixione:2008yi} we use the explicit form
\begin{equation}
d\sigma^{sub.}_{H^-t}=\frac{f_{BW}(m_{H^-\bar{b}})}{f_{BW}(m_{t})}\left|\tilde{{\cal A}}^{(t\bar{t})}\right|^2.
\label{DSsub}
\end{equation}
Here $f_{BW}$ is the Breit-Wigner function, and $\tilde{{\cal A}}^{(t\bar{t})}$ denotes the LO scattering amplitude for $tH^-b$ production including only doubly resonant diagrams (note that these form a gauge-invariant set), but with the kinematics reshuffled so as to place the $\bar{t}$ on-shell. We obtain the relevant matrix elements from MadGraph~\cite{Alwall:2007st,Maltoni:2002qb}. The use of the full amplitude ensures that spin correlations of the $\bar{t}$ are present in the decay products $H^-\bar{b}$ (these are not to be confused with the decay products of the top quark and charged Higgs boson, which do not currently have spin correlations included). Clearly this squared amplitude fulfills the first of the above requirements. It is then multiplied by a ratio of Breit-Wigner functions, which damp the subtraction term as $m_{H^-\bar{b}}$ moves away from the top mass, thus fulfilling the second condition above. More details can be found in~\cite{Frixione:2008yi}. 

Given the fact that the subtraction takes place at the cross-section level in DS, rather than at the amplitude level as in DR, the interference term between $H^-t$ and $t\bar{t}$ is still present in DS. It follows that the difference between DR and DS mostly measures the interference, as stated above. The difference between DR and DS is also affected by ambiguities in the formulation of the subtraction term, and by potential gauge invariance violation in DR. However, both of these effects are small~\cite{Frixione:2008yi}.

We will see in section~\ref{resultssmall} that the difference between DR and DS can indeed be made small subject to adequate cuts. Firstly, however, we present example results for the large Higgs mass region $m_{H^-}>m_t$, where the complications of this section are not relevant.
\section{Results}
\label{results}
In this section, we present example results from our MC@NLO implementation for $H^-t$ production. Our aim is not to undergo a thorough phenomenological analysis, but rather to present a few observables which demonstrate the differences between a fixed order description, and one that is matched to a parton shower. In the following results, we consider the LHC at 14 TeV, set renormalization and factorization scales to $\mu_F=\mu_R=(m_{H^-}+m_t)/2$, and use a top quark mass of $m_t=172$ GeV. For the charged Higgs Yukawa couplings we assume a type-II two-Higgs doublet model (eq.~(\ref{aandb})), with $\tan\beta=30$ and the NLO running bottom and top quark masses with scales as above. However, $\tan\beta$ is simply a scaling factor of the $bg \to t H^-$ cross section and once it lies above
$\tan\beta \gtrsim 10$ is has no impact on the relative size of the
NLO corrections or the distributions of the final state
particles~\cite{Zhu:2001nt,Plehn:2002vy,Berger:2003sm}. We use CTEQ5M1 partons throughout~\cite{Lai:1999wy}. Although more recent global parton analyses are available (e.g.~\cite{Tung:2006tb,Martin:2009iq}), this choice facilitates comparison with the results of~\cite{Plehn:2002vy,Berger:2003sm} where appropriate. 
\subsection{Large charged Higgs mass: $m_{H^-}>m_t$}
\label{resultslarge}
\begin{figure}
\includegraphics[width=0.48\textwidth]{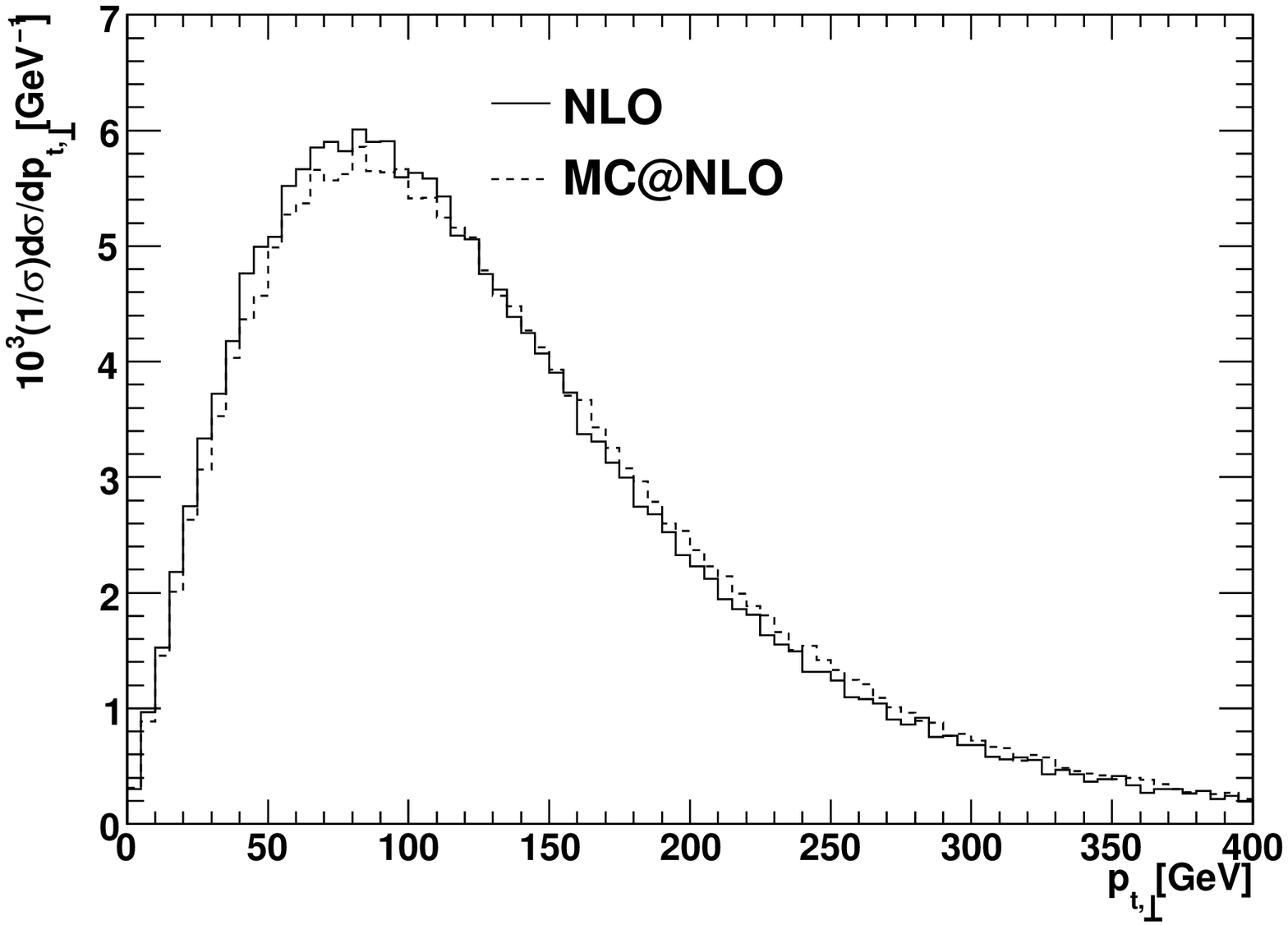}
\includegraphics[width=0.48\textwidth]{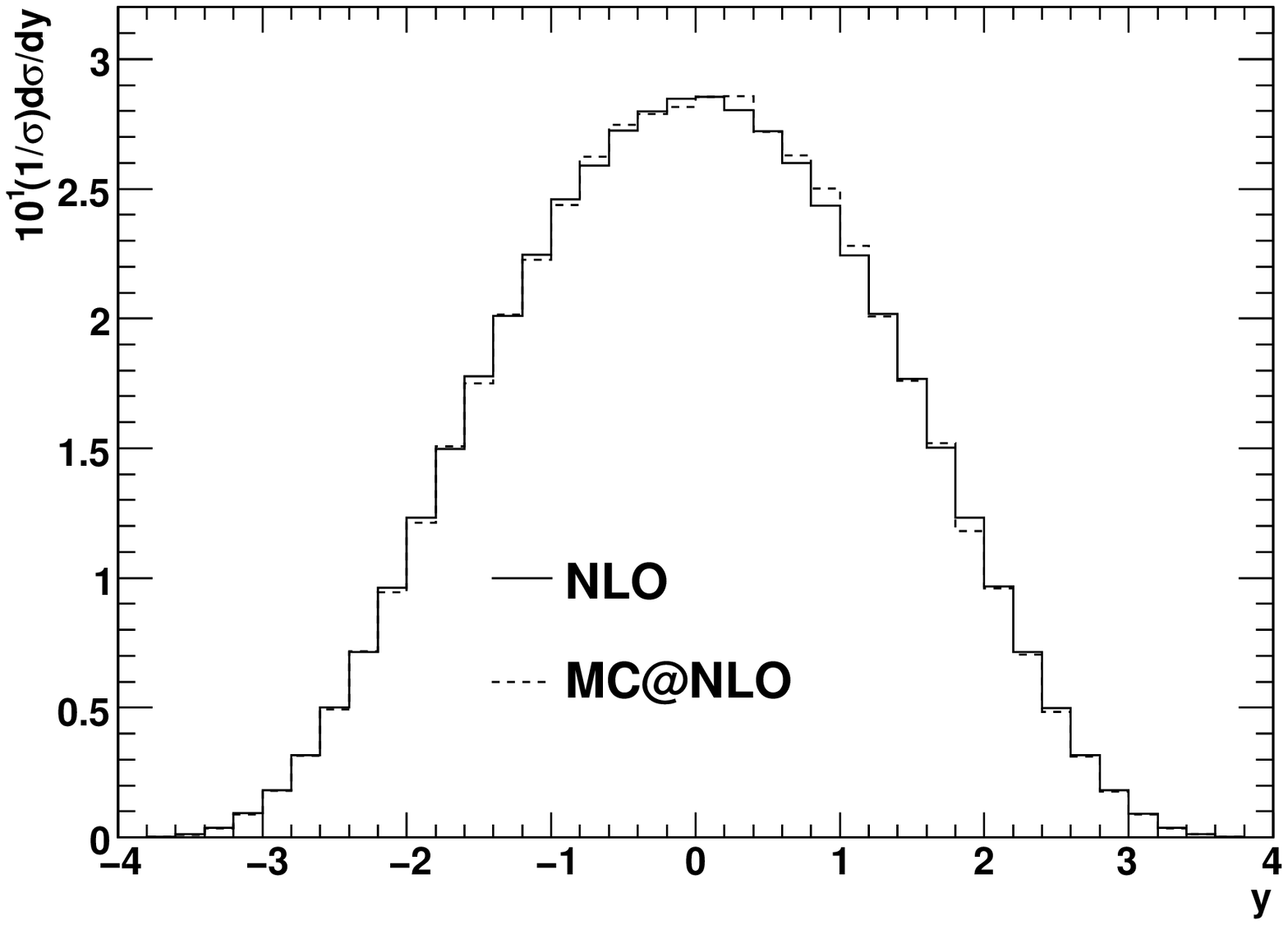}
\includegraphics[width=0.48\textwidth]{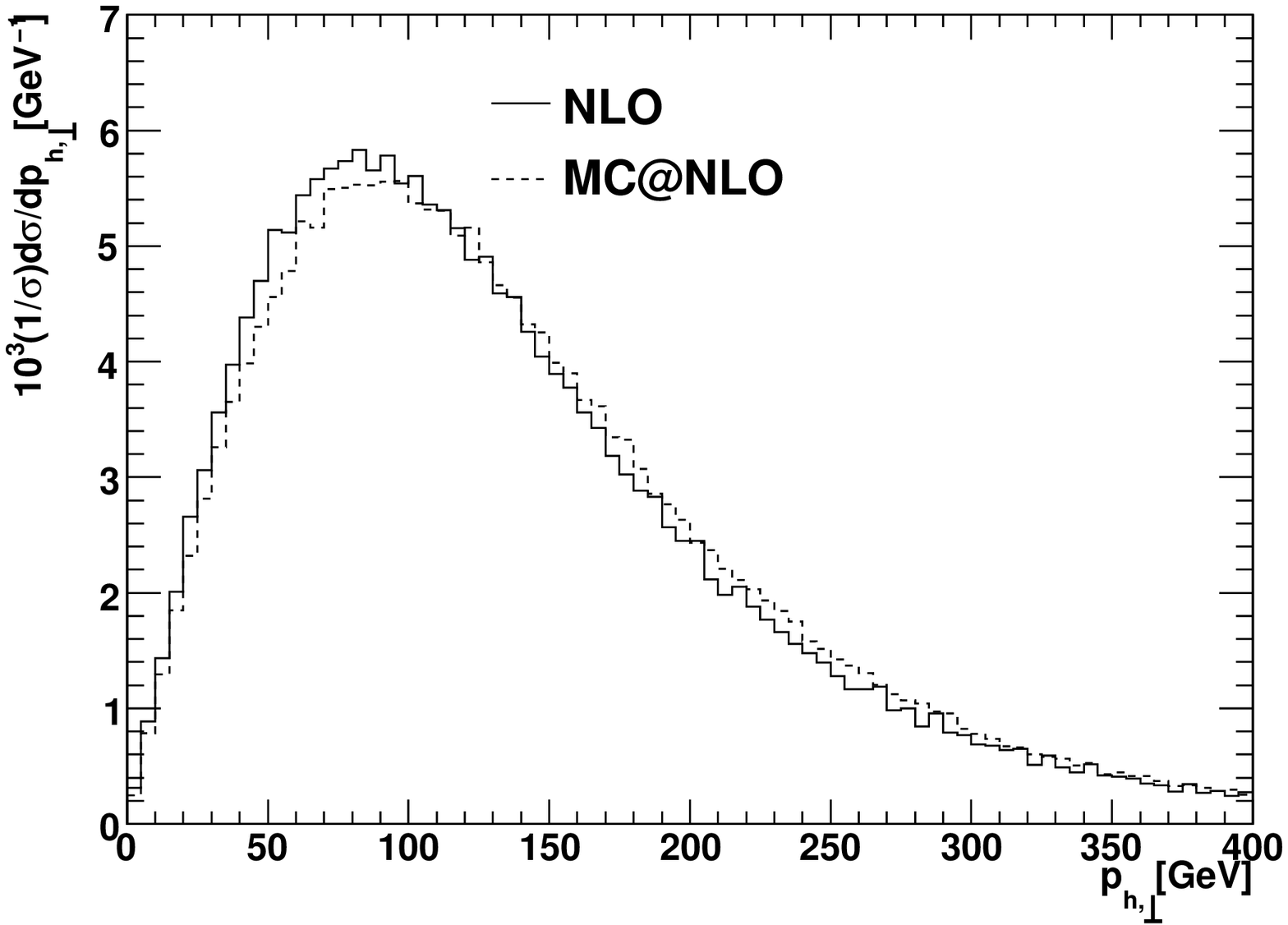}
\includegraphics[width=0.48\textwidth]{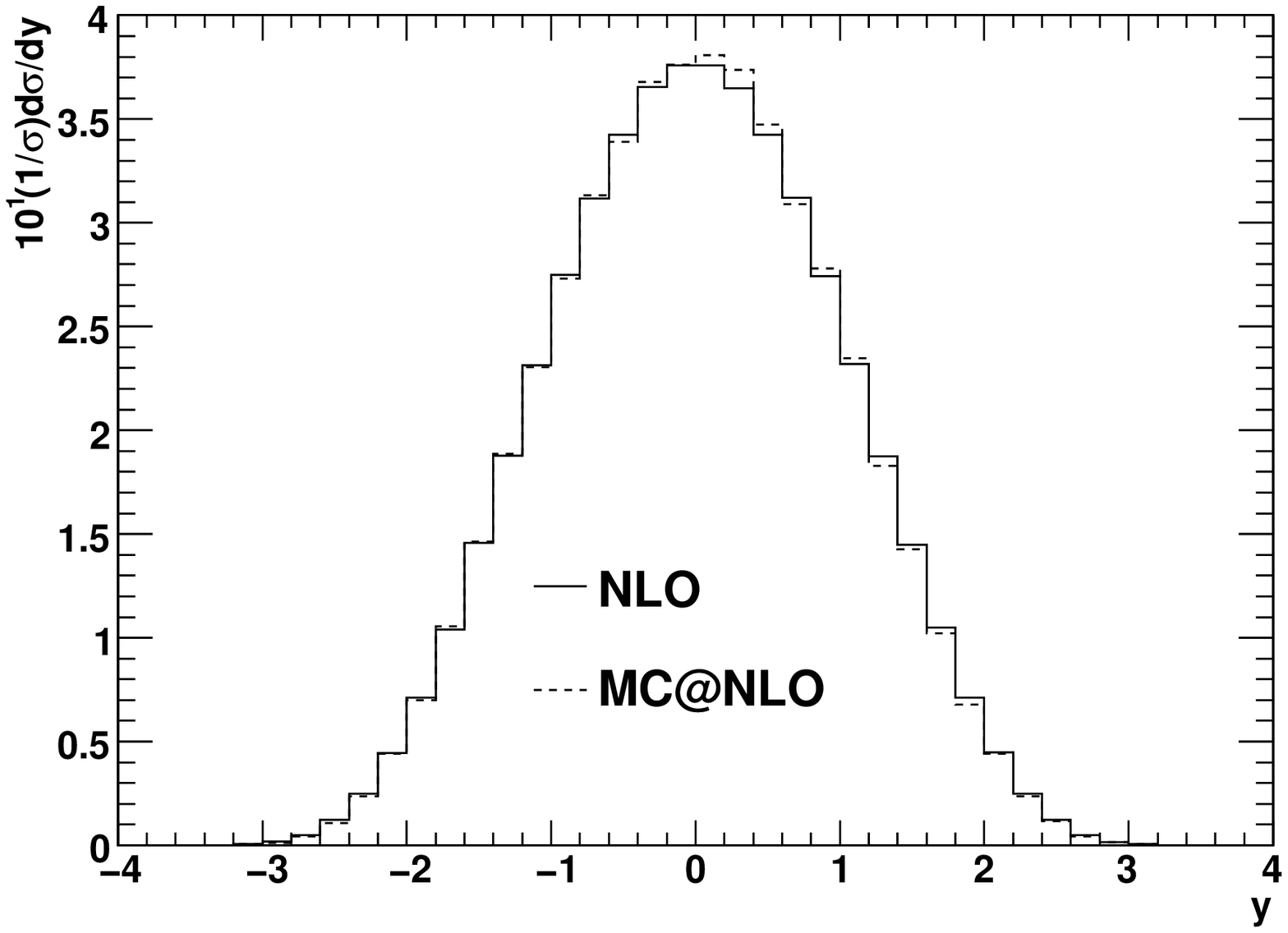}
\includegraphics[width=0.48\textwidth]{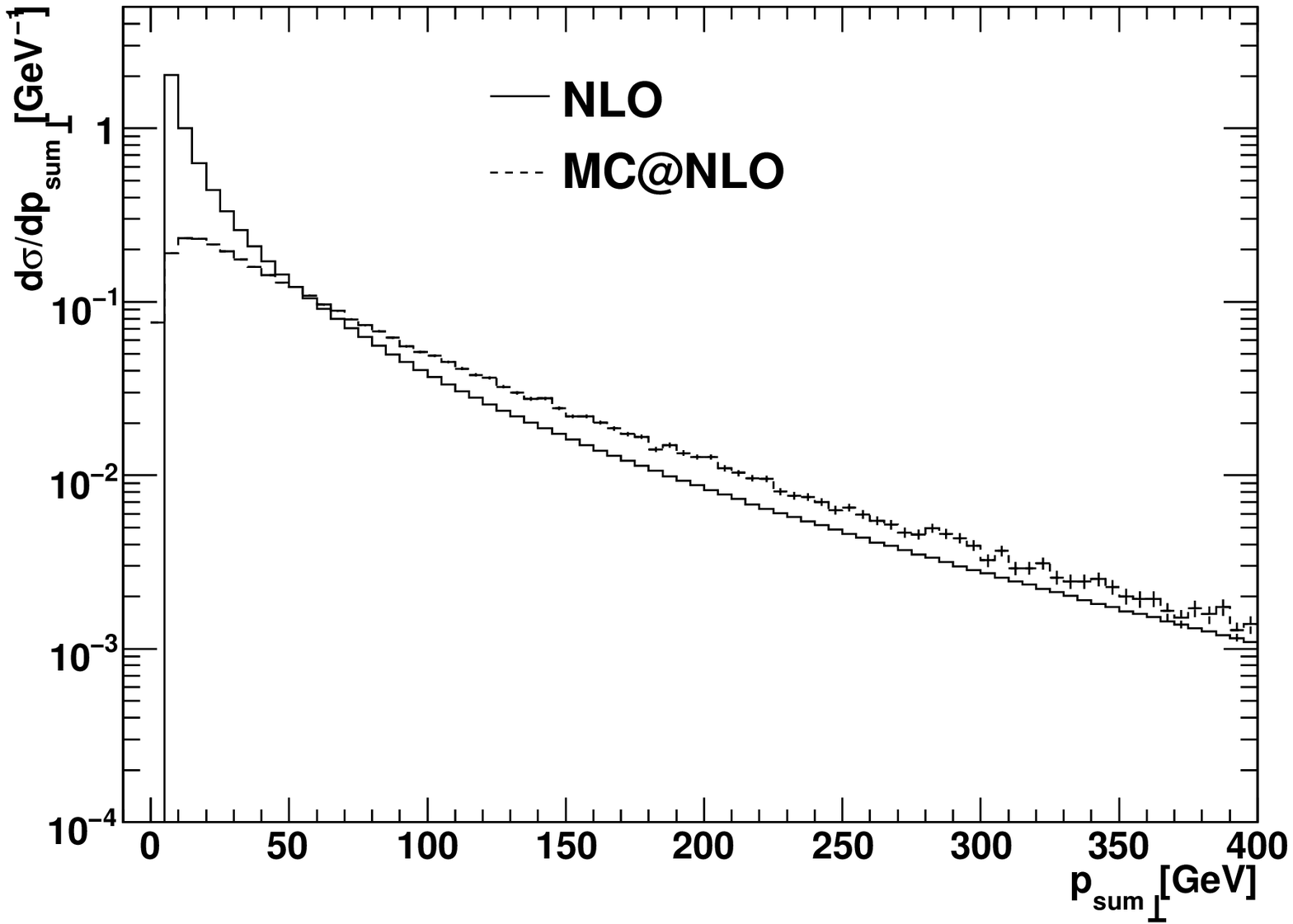}
\includegraphics[width=0.48\textwidth]{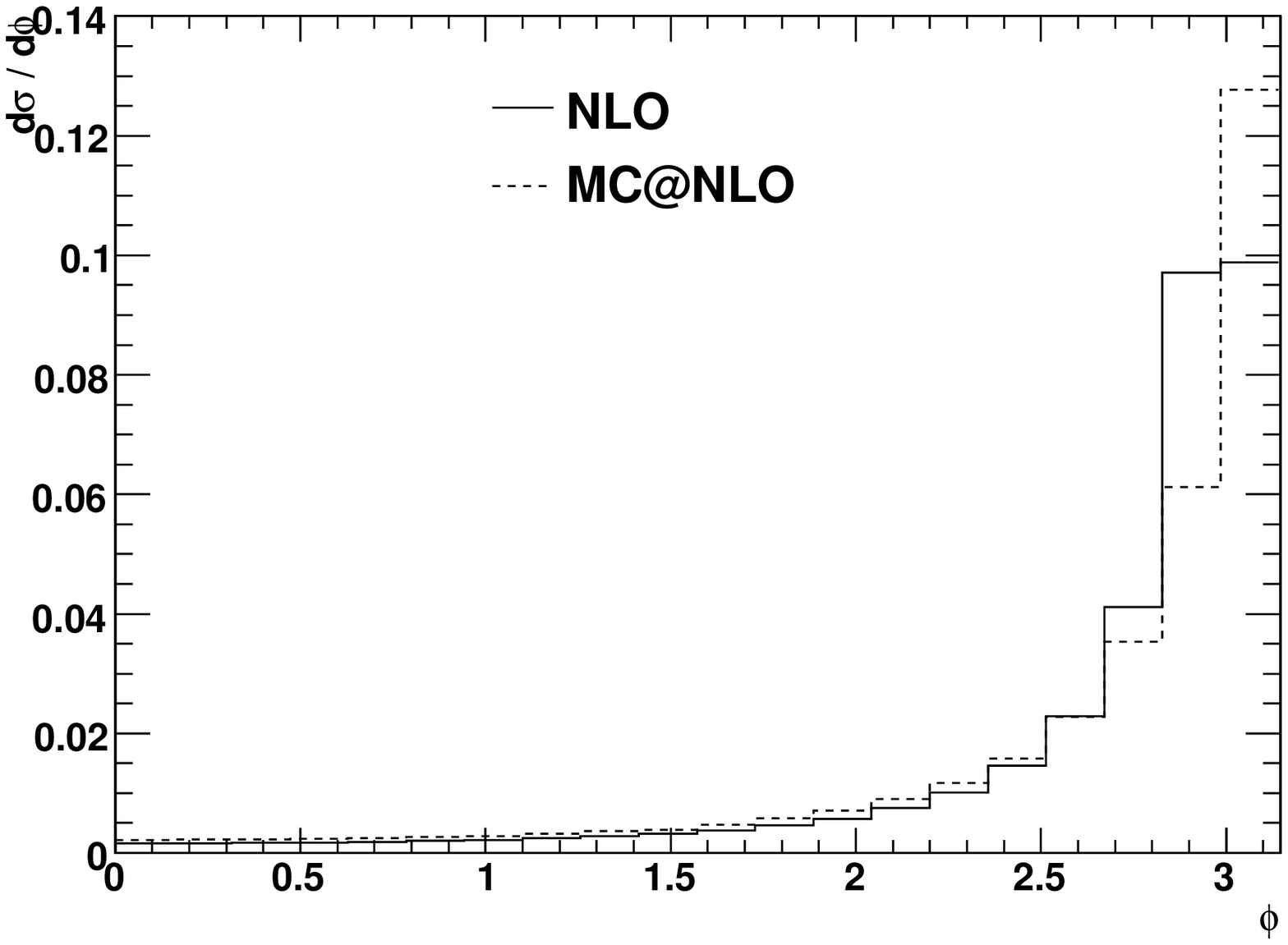}
\caption{Comparison of NLO and MC@NLO results, with parameters as given in the text. Shown are the transverse momentum and rapidity distributions of the top quark (upper line) and Higgs boson (middle line), the transverse momentum of the $H^-t$ system (bottom left) and the azimuthal angle between the top quark and charged Higgs boson (bottom right).}
\label{NLOfigs}
\end{figure}

In this section we consider, unless otherwise stated, $m_{H^-}$=300 GeV. We begin by showing the transverse momentum and rapidity distributions of the top quark and charged Higgs boson in figure~\ref{NLOfigs}.
One sees that there is not a great deal of difference between NLO and MC@NLO descriptions of these observables, as perhaps expected given the inclusive nature of these observables, and the fact that they are not subject to logarithmic corrections. Instead, the $p_t$ and rapidity distributions act as a consistency check between the NLO and MC@NLO calculations.

We now consider observables designed to manifest the differences between the fixed order and parton showered approaches. The quantity
\begin{equation}
p_{t,sum}=|\vec{p}_{t,H}+\vec{p}_{t,top}|
\label{ptsumdef}
\end{equation}
(i.e. the magnitude of the vector sum of transverse momentum of the $H^-t$ system) is shown in figure~\ref{NLOfigs} (bottom left). At LO, this distribution would be a Dirac delta function at $p_{t,sum}=0$ due to 4-momentum conservation. At NLO, there is a sharp rise as $p_{t,sum}\rightarrow0$ due to the real emission contributions, whilst the zero bin is negative due to virtual corrections. This behavior is smoothed by the parton shower, and the MC@NLO curve indeed displays the characteristic Sudakov behavior.

Also of interest is the distribution of the azimuthal angle $\phi$ between the top quark and charged Higgs, shown in figure~\ref{NLOfigs} (bottom right). At LO, this would be a delta function at $\phi=\pi$ i.e. the final state particles are produced back-to-back. At NLO, this gets decorrelated by the emission of one hard parton. There is also a suppression of the $\phi=\pi$ bin due to the virtual corrections, resulting in a slightly peculiar shape in the last two bins. When the full parton shower is added this feature is smoothed out to produce a more physical-looking decorrelation, thus demonstrating the advantages of the parton shower approach.

The results of this section indicate that the MC@NLO is indeed working as expected. We now consider some phenomenological properties of $b$ and light jets. Our motivation is as follows. Previous analyses of the $H^{-}t$ process (such as~\cite{Barnett:1987jw,Bawa:1989pc,Barger:1993th}) have argued that additional $b$ jets, by which we mean $b$ jets that are not the hardest $b$ jet and thus are less likely to have come from the decay of the top quark, have different kinematic properties (specifically the transverse momentum distribution) to light jet radiation. Furthermore, that one may exploit this difference to design event selection criteria, i.e. reduce backgrounds. A typical example is the background due to vector boson plus multijet (or pure multijet) production. In a non-negligible fraction of cases, light jets in the multijet background may be mistagged as $b$ jets, and therefore be mistaken for signal events. It is much less likely, however, that two light jets will be mistagged, such that if one can convincingly identify events with two $b$ jets, the signal to background ratio may be significantly enhanced. 

This naturally raises two questions. First of all, what are the fractions of events in which radiated light jets or additional $b$ jets are present? Secondly, are additional $b$ jets harder than radiated light jets? An MC@NLO description is better suited to answering this question than a purely fixed order description, owing to the greater number of final state partons present, and we therefore consider these questions here.

In the following we will consider cases where the top quark decays both leptonically, and also hadronically. The former case is more promising experimentally, due to the absence of hard jets from the top decay (we will see this in more detail in what follows). We will assume, unless otherwise stated, 100\% $b$-tagging efficiency. Furthermore, we will not consider systematic uncertainties due to reconstruction of the top quark or charged Higgs kinematics (i.e. we will assume that the decay products are sufficiently hard in $p_t$ to be observable). We cluster jets according to a $k_t$ algorithm with $D=0.7$, and require all jets to lie in the $b$-tagging detector volume
\begin{equation}
|\eta|<2.5 \qquad \qquad p_T>25~\text{GeV}.
\label{eq:detector}
\end{equation}

\begin{figure}
\includegraphics[width=0.48\textwidth]{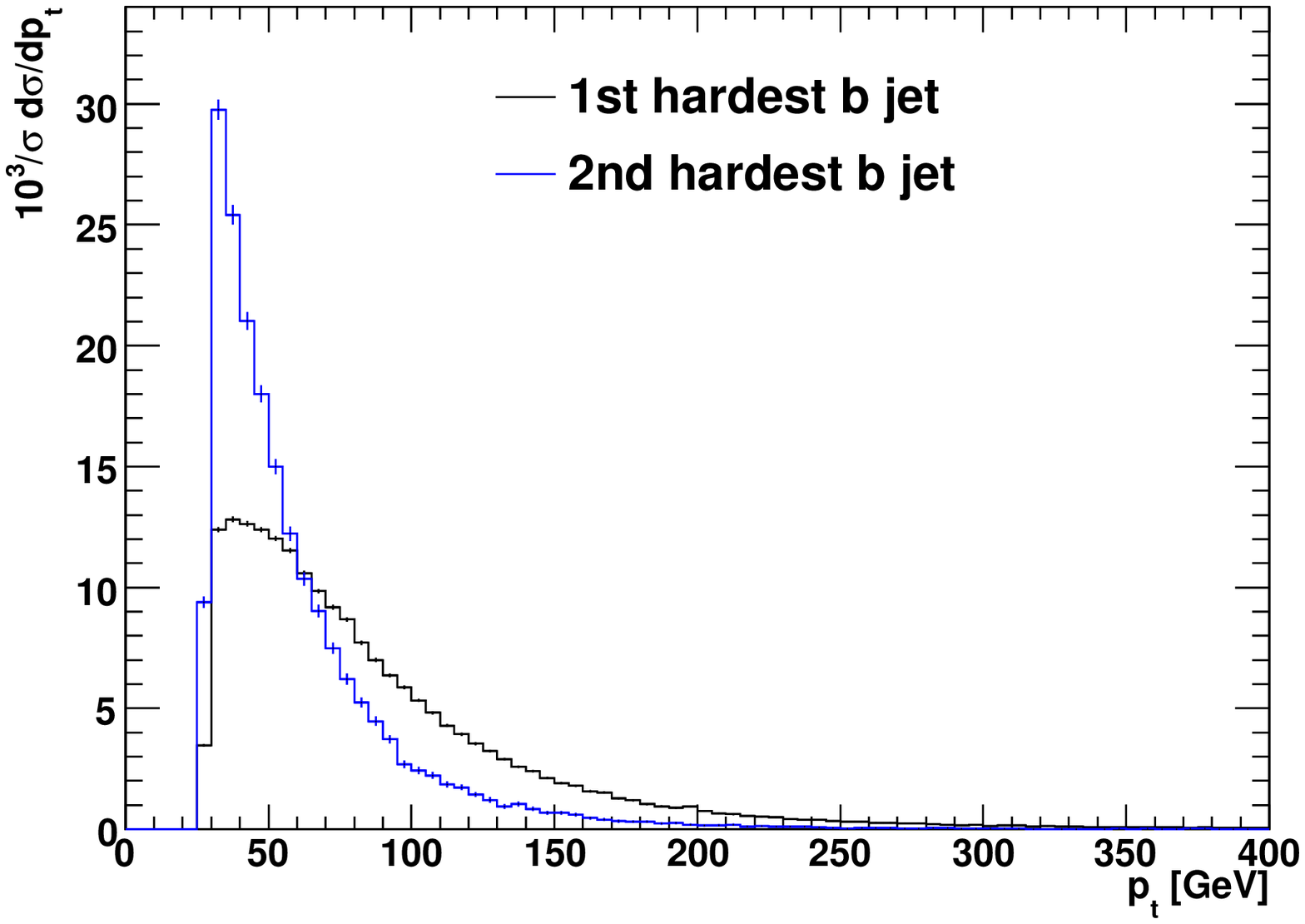}
\includegraphics[width=0.48\textwidth]{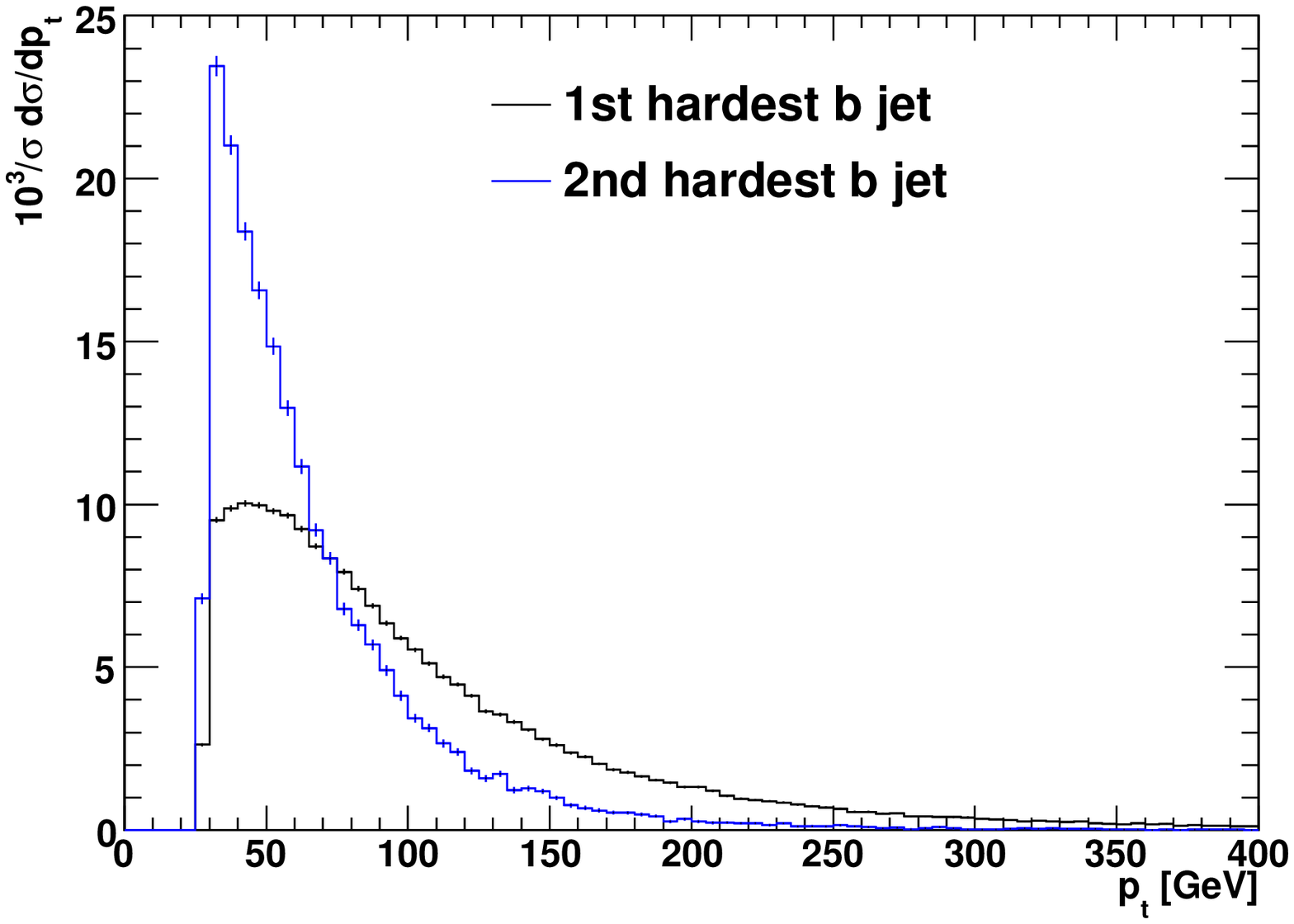}\\
\includegraphics[width=0.48\textwidth]{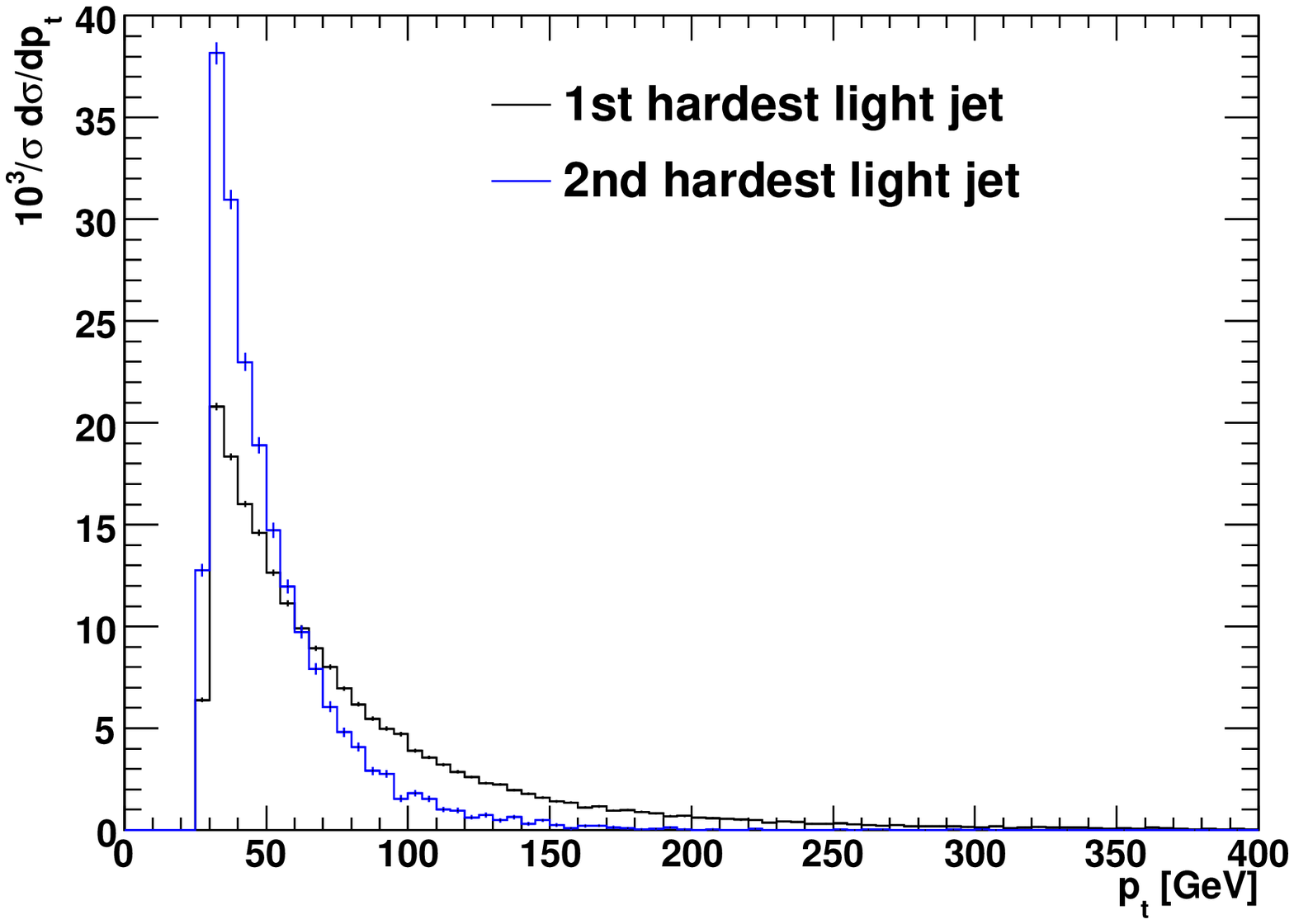}
\includegraphics[width=0.48\textwidth]{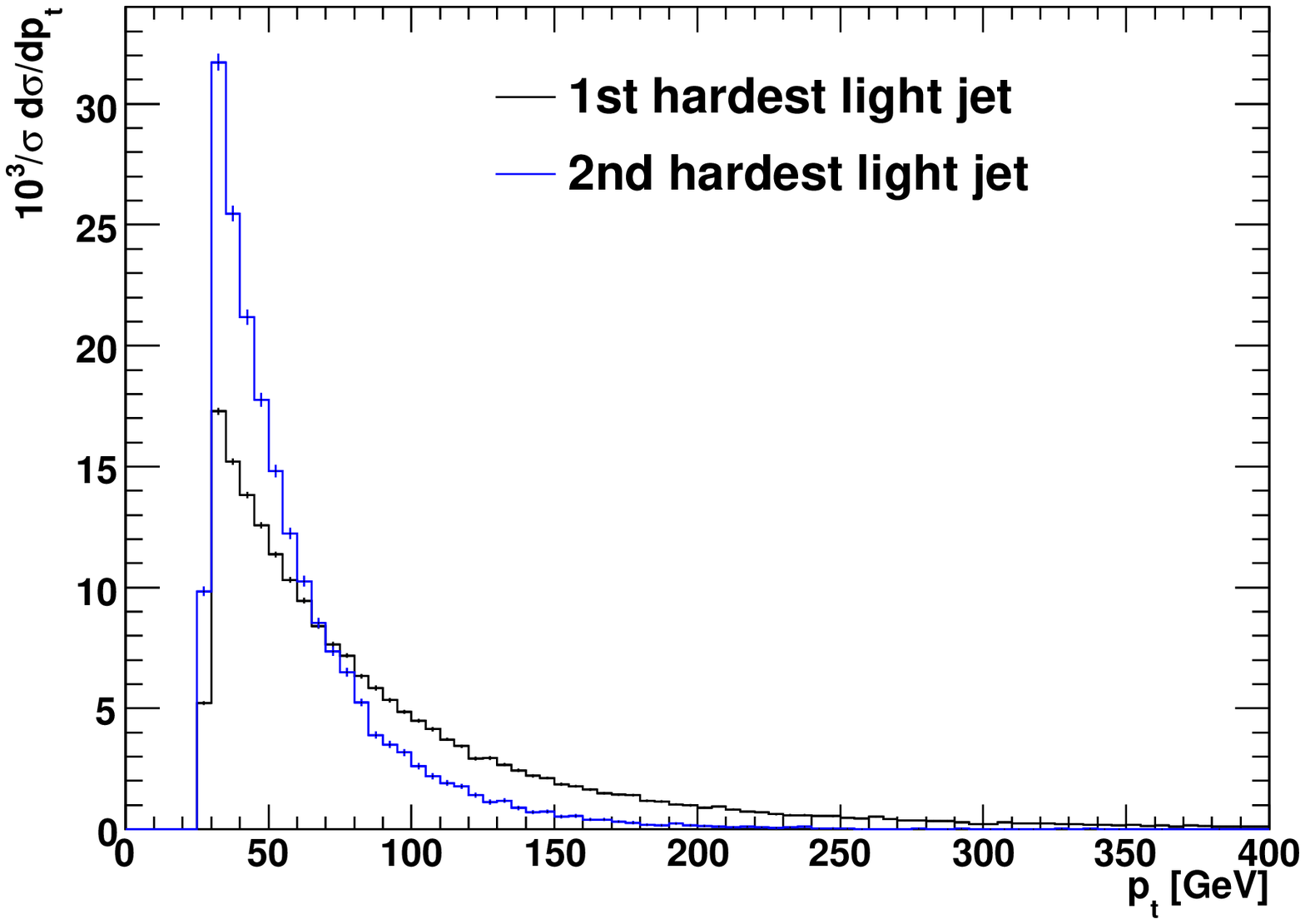}\\
\includegraphics[width=0.48\textwidth]{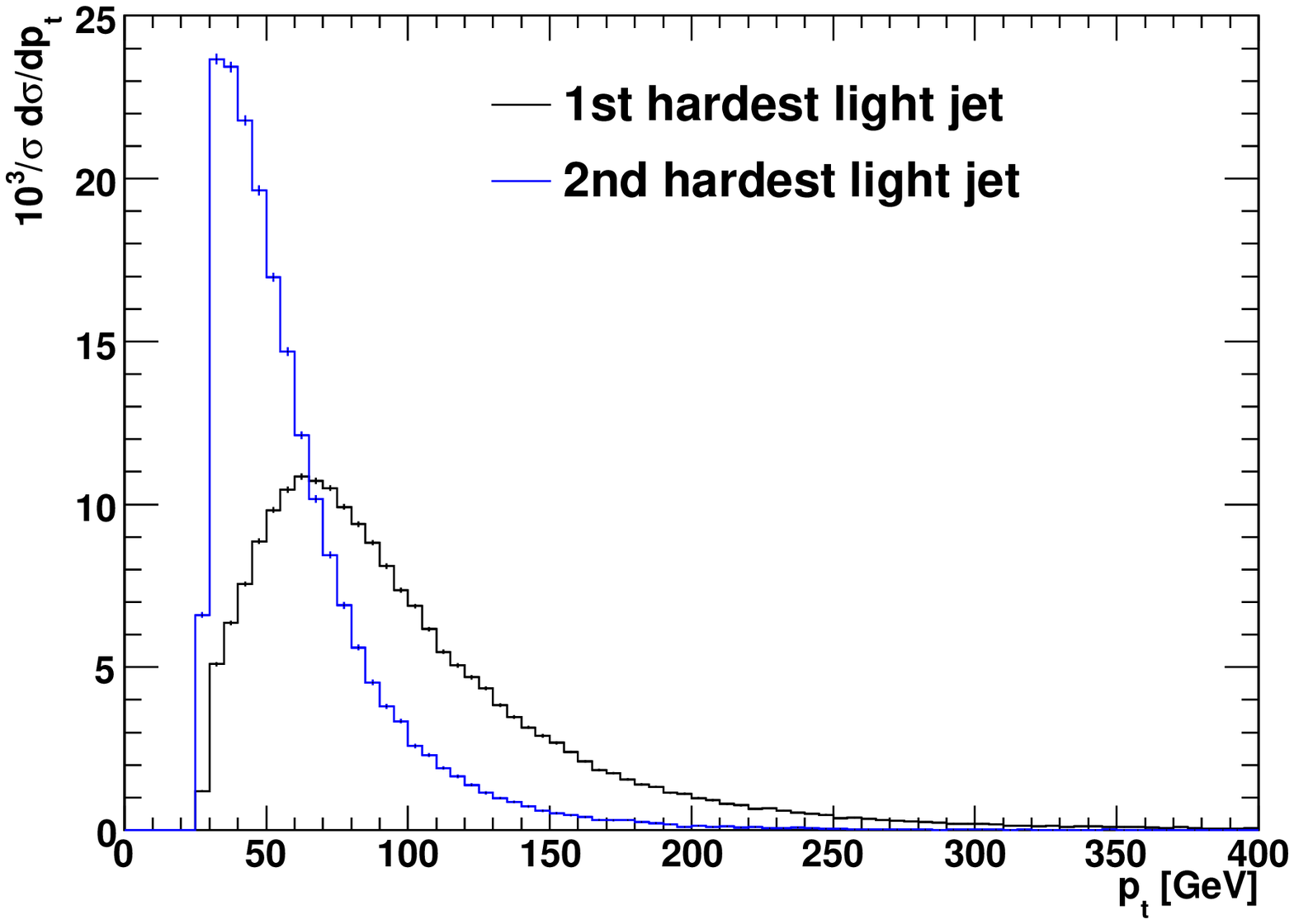}
\includegraphics[width=0.48\textwidth]{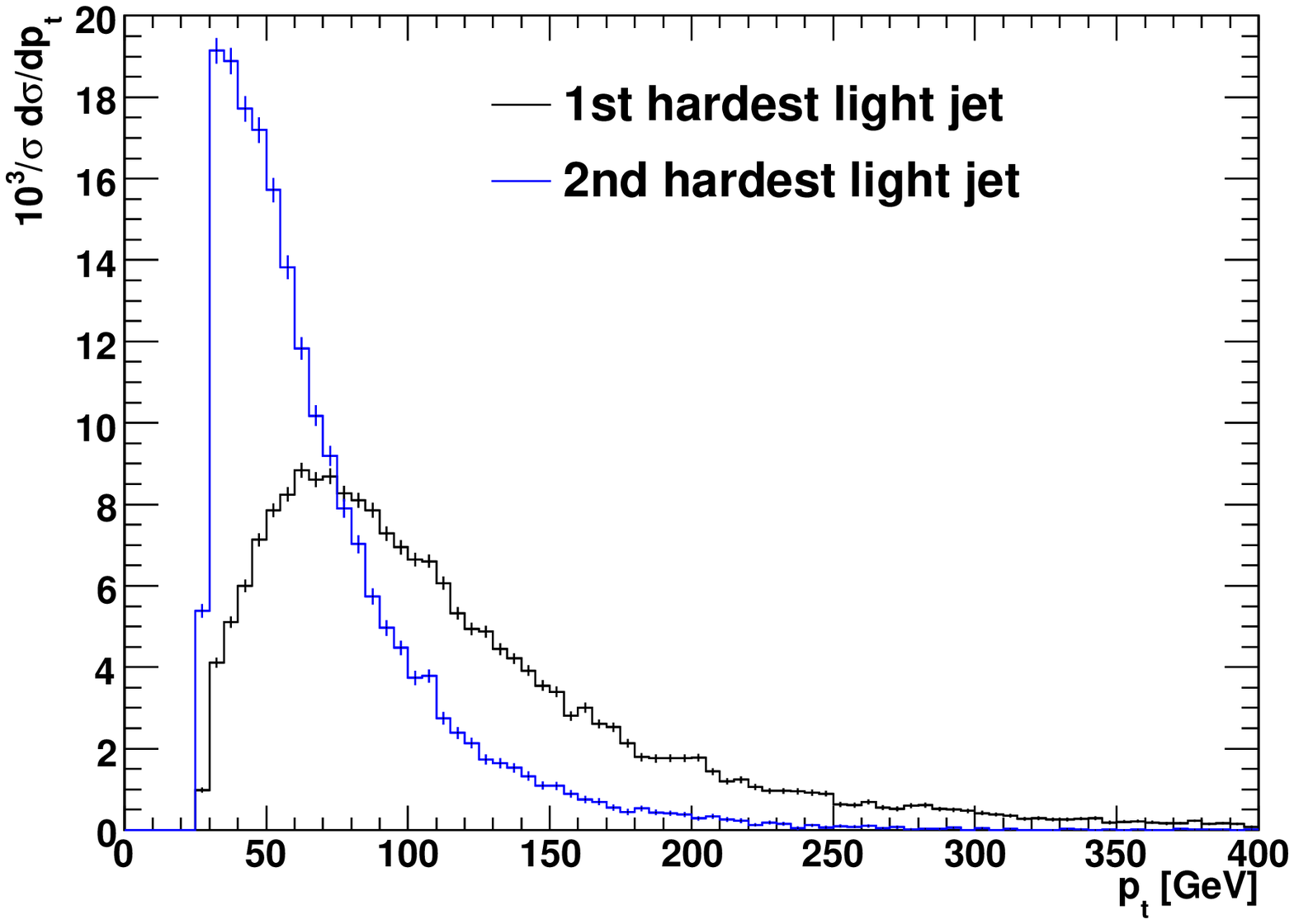}\\
\caption{Normalized transverse momentum distributions of
  the two hardest jets. We include the detector cuts of
  eq.~(\ref{eq:detector}). First row: the two hardest $b$ jets; second
  row: the two hardest light-flavor jets for a leptonic top decay;
  third row: the two hardest light-flavor jets for a hadronic top
  decay. The left-hand column corresponds to $\mh =300$~GeV, and the
  right-hand column to $\mh =800$~GeV.}
\label{fig:jet_distri_pt}
\end{figure}

\begin{figure}
\includegraphics[width=0.48\textwidth]{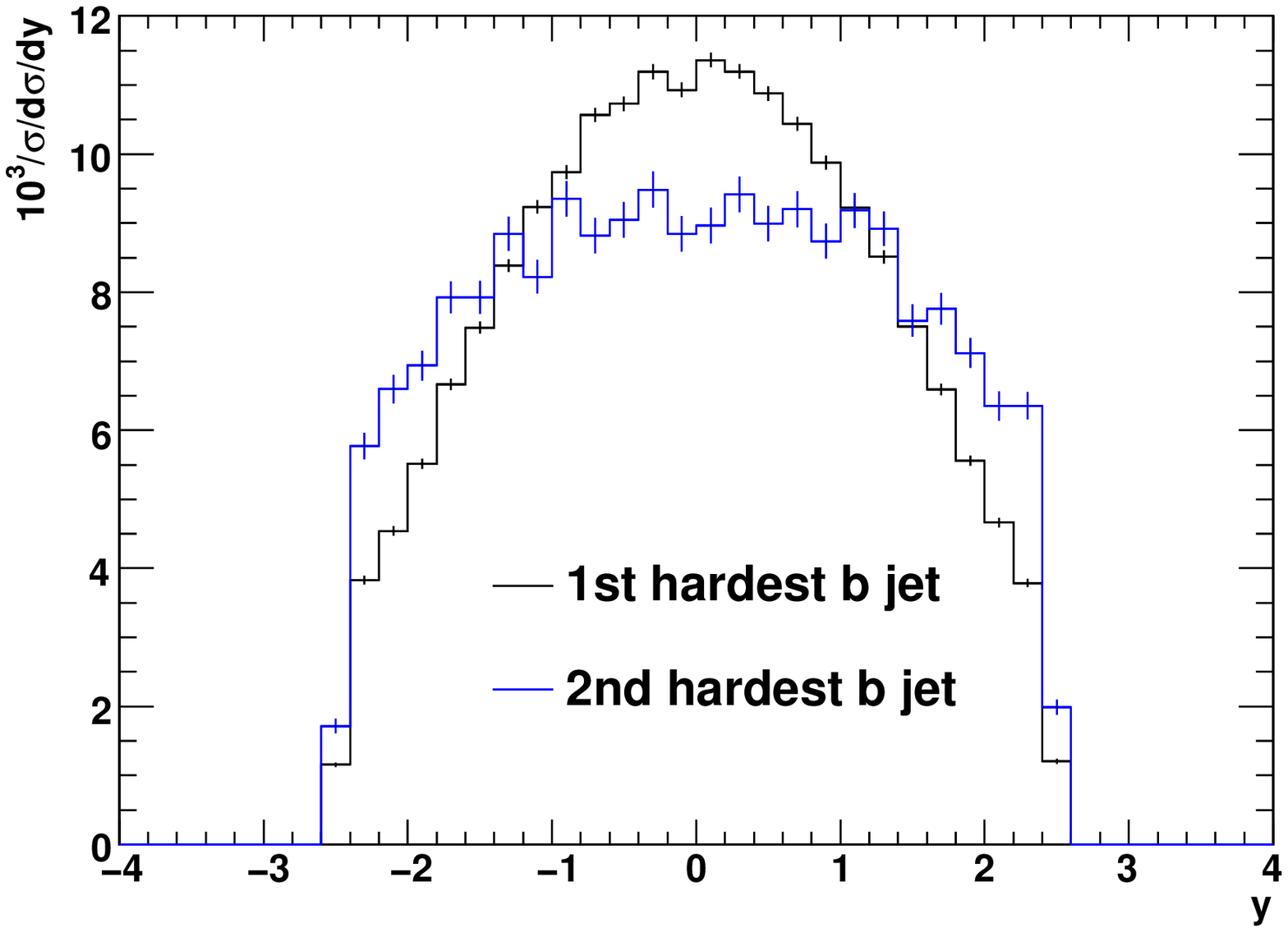}
\includegraphics[width=0.48\textwidth]{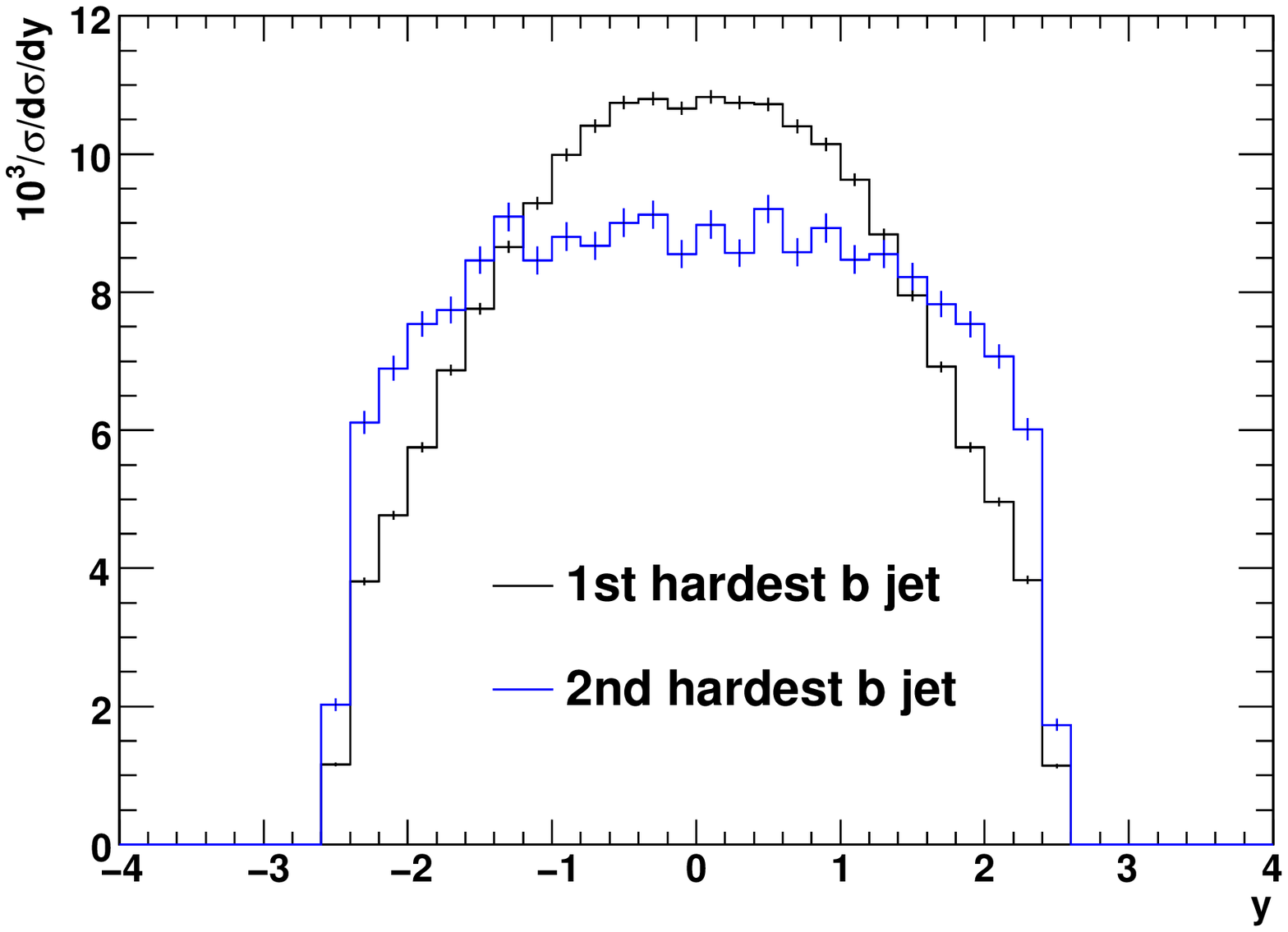} \\
\includegraphics[width=0.48\textwidth]{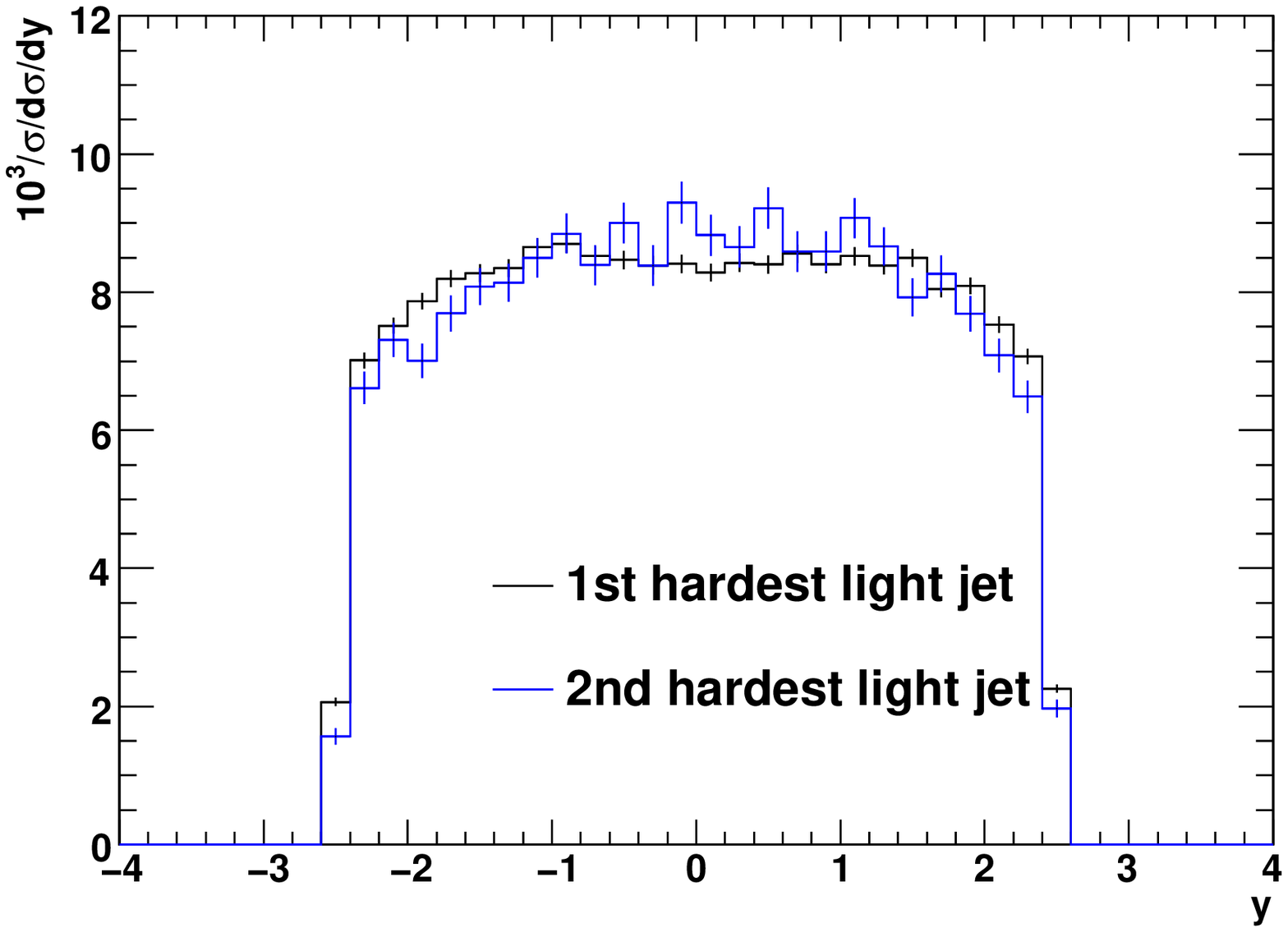}
\includegraphics[width=0.48\textwidth]{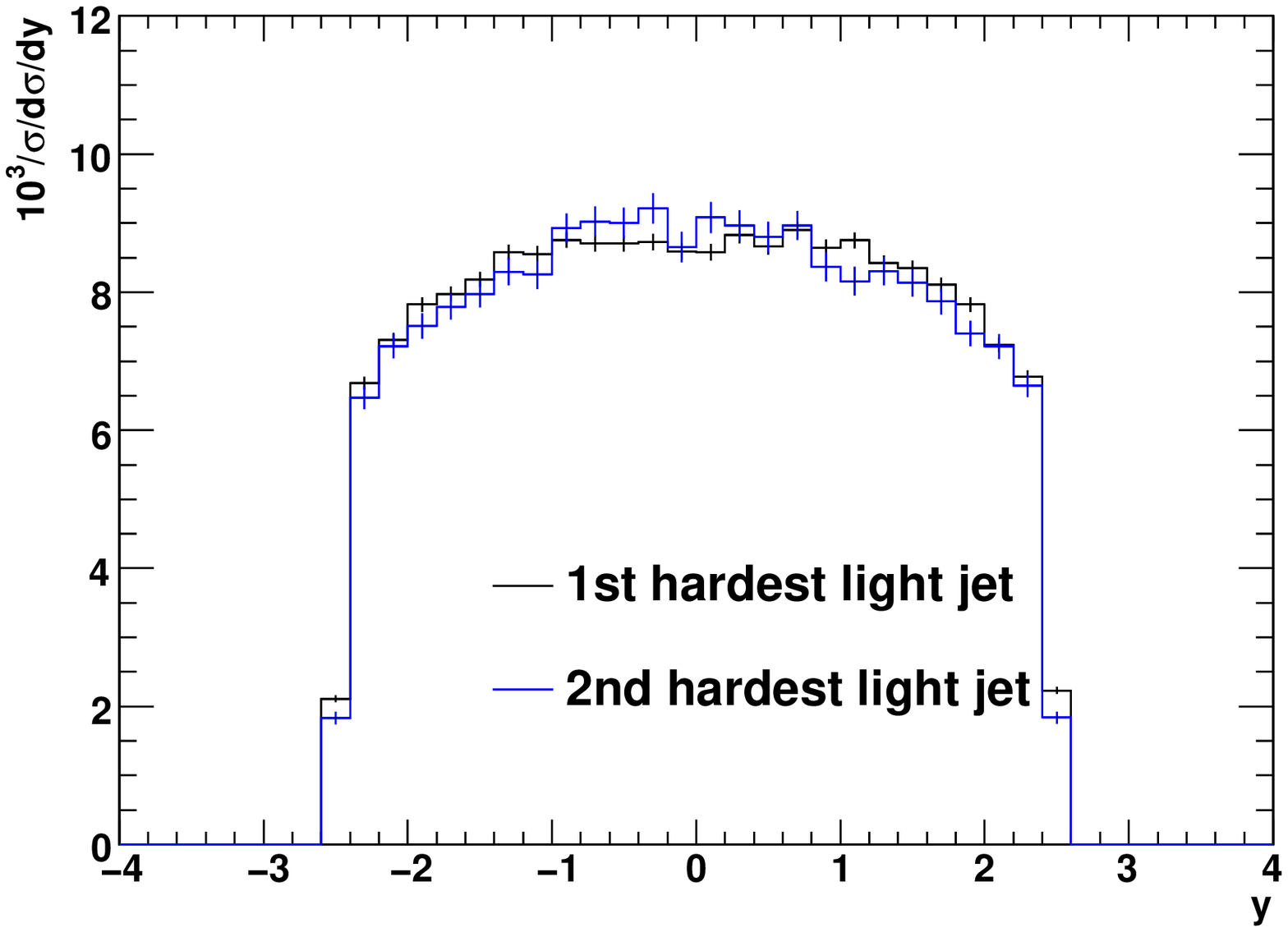} \\
\includegraphics[width=0.48\textwidth]{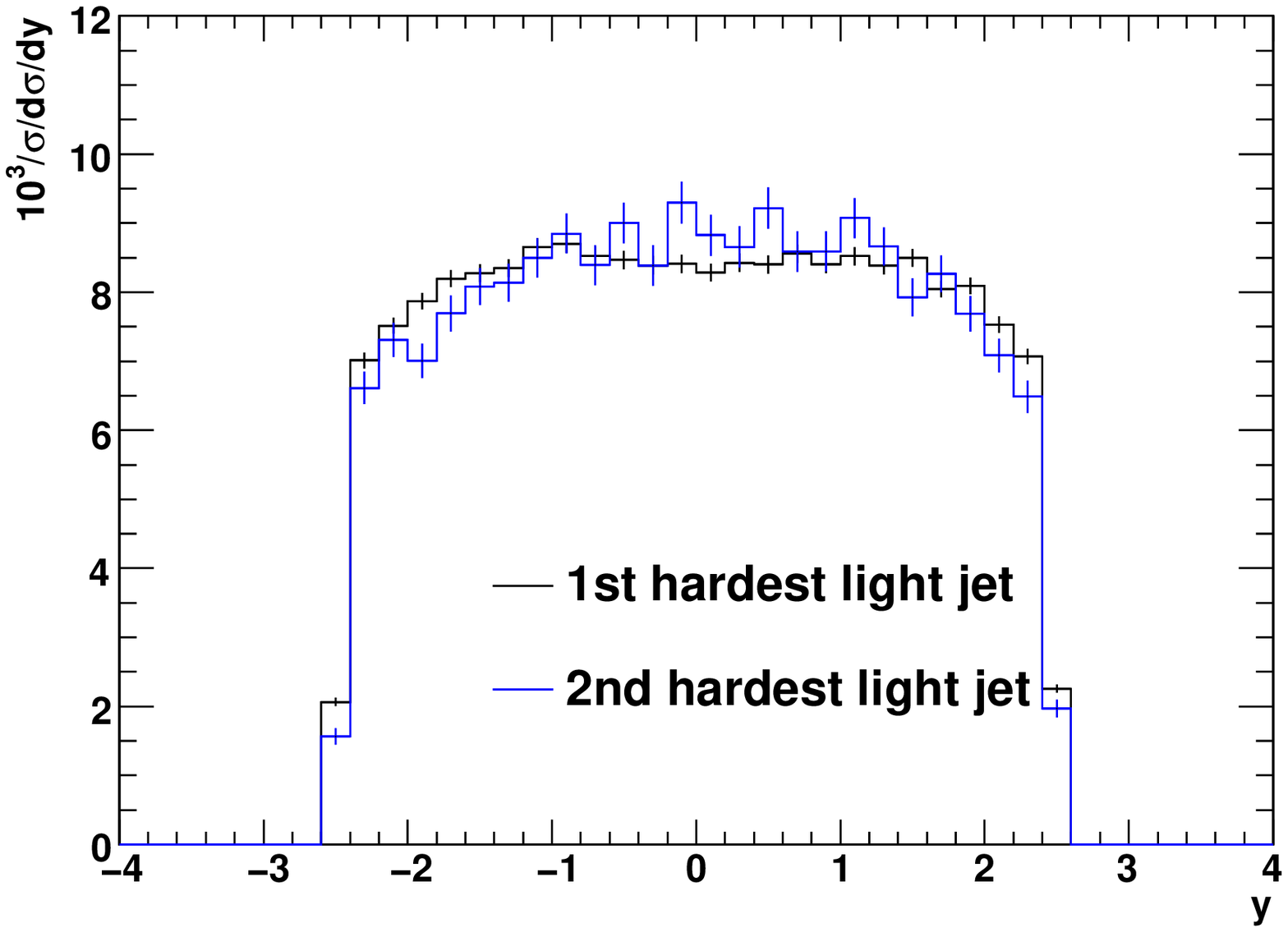}
\includegraphics[width=0.48\textwidth]{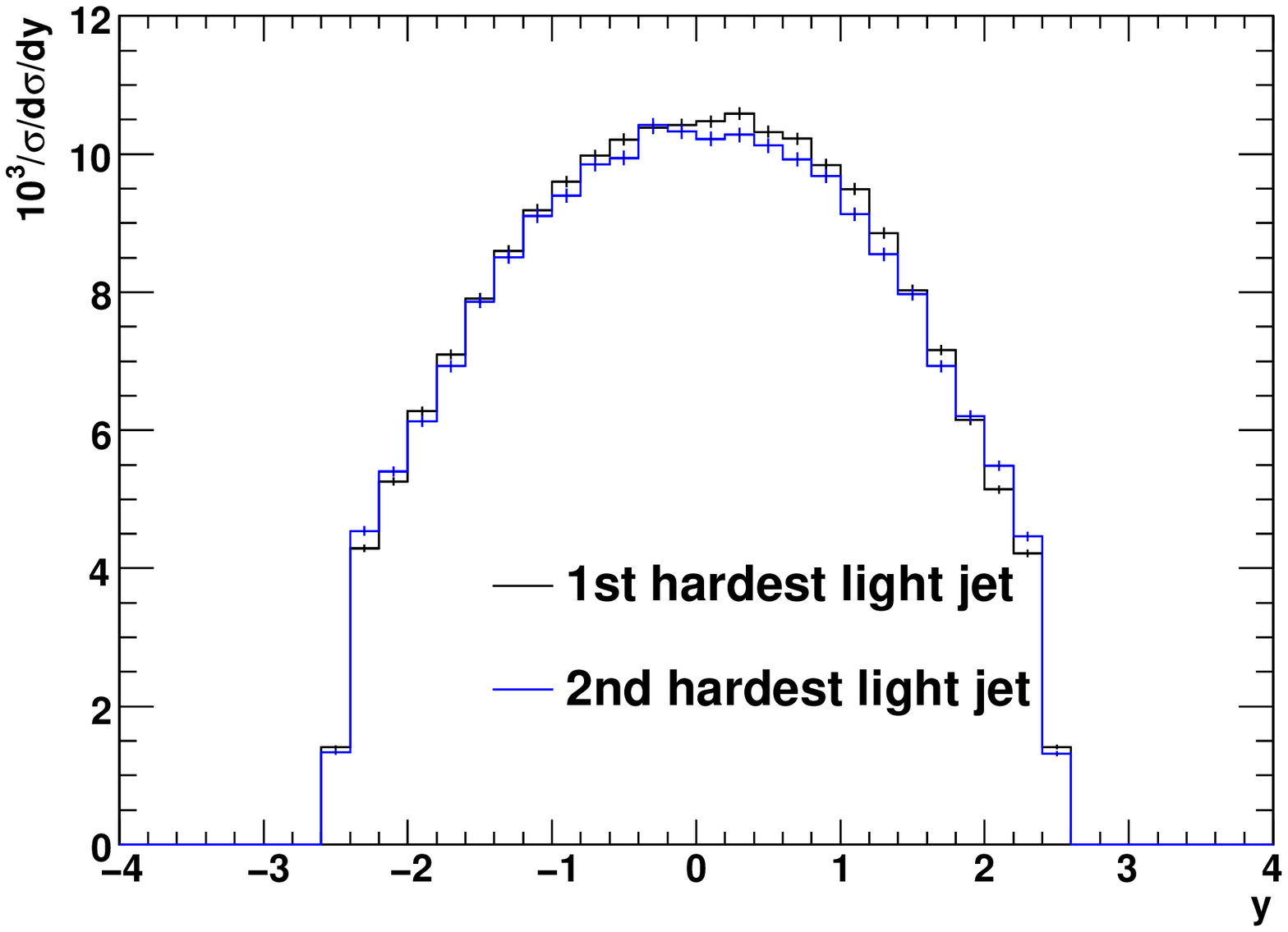} \\
\caption{Normalized rapidity distributions of
  the two hardest jets. We include the detector cuts of
  eq.~(\ref{eq:detector}). First row: the two hardest $b$ jets; second
  row: the two hardest light-flavor jets for a leptonic top decay;
  third row: the two hardest light-flavor jets for a hadronic top
  decay. The left-hand column corresponds to $\mh =300$~GeV, and the
  right-hand column to $\mh =800$~GeV.  }
\label{fig:jet_distri_y}
\end{figure}

In any given event, one sees a number (possibly zero) of $b$ jets, but has no way of knowing which $b$ jet came from the decay of the top quark. Thus, in comparing the properties of additional $b$ jets with radiated light jets, one assumes that the hardest $b$ jet in any event arises from the top quark, and that the second hardest $b$ jet arises from QCD radiation (in our calculation this happens either in the NLO matrix element, or in the parton shower; likewise for light jets that do not come from a top decay). 
Figures~\ref{fig:jet_distri_pt} and~\ref{fig:jet_distri_y} show the transverse momentum and rapidity
distributions of the two hardest bottom and light-flavor jets, if
present, for two charged Higgs masses 300~GeV and 800~GeV. The light jet results are shown for leptonic as well as hadronic top decays. In the
first row of distributions we see the behavior of the two bottom jets,
one from the top decay and the other (mostly) from initial-state gluon
splitting. The harder of the two bottom jets peaks at transverse
momenta around 50~GeV, as indeed is expected from the three-body decay of a top quark.
The position of this peak therefore does not change with the
Higgs mass. The softer of the two bottom jets is peaked at small
transverse momenta and extends to larger rapidities than the decay jet
--- the typical pattern of initial state radiation. Thus, the assumption that the first hardest jet arises from the top decay and the second hardest from QCD radiation appears to be well-founded.
At the higher Higgs mass (i.e. a larger scale in the hard process) both rapidity
distributions become flatter. This can happen because of enhanced collinear radiation in the presence of a harder QCD scale, or because of momentum conservation effects between the harder Higgs boson and its accompanying jets.  

For leptonic top decays all light-flavor jets arise from QCD
radiation, rather than from decay of the top quark. This is reflected in the sharp drop of the $p_T$
distributions as well as the flat $\eta$ behavior. As mentioned above, it is interesting to compare the properties of the second hardest $b$ jet (i.e. the $b$ jet mostly coming from QCD radiation rather than the top quark decay) with the hardest light jet. In particular, one may compare the transverse momentum distributions shown in figure~\ref{fig:jet_distri_pt}. One sees that the difference between the second hardest $b$ jet and hardest light jet distributions is not large. If
anything, for $\mh = 300$~GeV the $p_T$ distribution of the
hardest light-flavor jet tends to fall off slower than the
second hardest bottom jet distribution. This behavior is more clearly visible
for heavy Higgs masses, where the $p_T$ distribution reaches half its
maximum value at 55~GeV for the second bottom jets and at 65~GeV for
the hardest light-flavor jet. In principle, both the light-flavor jet
and the bottom jet are radiated collinearly, so the source of this
slight discrepancy is simply that there are more light-flavor jets to
pick from.

One may also consider hadronic decays, and results for the transverse momentum and rapidity distributions of the two hardest light jets are shown in the lower lines of figures~\ref{fig:jet_distri_pt} and~\ref{fig:jet_distri_y}. One sees that the hardest jets become significantly harder and more central, as expected from the fact that they now come predominantly from the top decay rather than QCD radiation. Unfortunately, this distinguishing feature is limited in its
experimental use, because the high mass scales in this process (stemming from the top quark and Higgs boson) allow for initial and final state radiation comparable in hardness to the jets from the top decay~\cite{Plehn:2005cq,Plehn:2008ae,Alwall:2008qv}. Indeed $W$ bosons and top quarks that undergo hadronic decays can only be reconstructed in special kinematic regions, such as when they are highly boosted~\cite{Butterworth:2002tt,Thaler:2008ju,Kaplan:2008ie,Almeida:2008yp,Plehn:2009rk}.



\begin{table}[t]
\begin{small} \begin{center}
\begin{tabular}{|cc||c|c|c|c|c||c|c|c|c|c|}
\hline
&&\multicolumn{5}{c||}{$\mh = 300$~GeV}
 &\multicolumn{5}{c|}{$\mh = 800$~GeV}\\
&&\multicolumn{5}{c||}{$\eta_{\text{cut}}$}
 &\multicolumn{5}{c|}{$\eta_{\text{cut}}$}\\
& $p_{T,\text{cut}}$
& 2.5 & 2.0 & 1.5 & 1.0 & 0.5
& 2.5 & 2.0 & 1.5 & 1.0 & 0.5\\
\hline
\hline
\multirow{4}{*}{$(a)$}
& 25~\gev &45.9  &40.0  &32.7  &23.9  &13.0 
     & 54.8 &48.8  &41.0  &31.0  &17.9 \\
& 45~\gev & 32.4 & 27.8 &22.3  &16.1  &9.0 
     & 41.7 &36.7  &30.5  &23.0  &13.7 \\
& 65~\gev &22.3  &18.8  &14.7  &10.4  &5.8 
     &30.9  &27.0  &22.2  & 16.5 &10.2 \\
& 85~\gev &16.2  &13.4  &10.3  &7.3  &4.2 
     & 23.6 &20.5  &16.6  &12.1  &7.4 \\
\hline
\hline
\multirow{4}{*}{$(b)$}
& 25~\gev &94.9 &91.0 &84.3  &72.2  &48.4 
     &95.8  &92.5  &86.3  &75.0  &52.0 \\
& 45~\gev &83.2  &79.2  &72.3  &61.0  &39.9 
     &87.1  &83.3  &76.8  &65.7  &45.2 \\
& 65~\gev &60.9  &57.3  &51.7  &43.2  &28.8 
     &70.5  &66.9  &61.3  &51.9  &35.9 \\
& 85~\gev &44.4  &41.5  &37.1  &31.1  &21.3 
     &56.2  &53.3  &48.6  &41.0  &28.7 \\
\hline
\hline
\multirow{4}{*}{$(c)$}
& 25~\gev &17.8  &14.3  &10.0  &5.7  &2.3 
     & 21.4 &17.1  &12.1  &7.1  &3.0 \\
& 45~\gev & 12.9 &10.6  &7.6  &4.5  &1.8 
     &16.6  &13.7  &9.9  &6.0  &2.5 \\
& 65~\gev &9.4  &8.0  &5.9  &3.5  &1.6 
     &12.3  &10.5  &7.9  &4.9  &2.0 \\
& 85~\gev & 7.2 &6.4  &4.8  &3.0  &1.4 
     &9.7  &8.5  &6.5  &4.1  &1.7 \\
\hline
\end{tabular}
\end{center} \end{small}
\caption{Probability (\%) to see (a) a light jet given that a $b$ jet has been observed, where the top decays leptonically; (b) a light jet given that a $b$ jet has been observed, where the top decays hadronically; (c) an additional $b$ jet given that one $b$ jet has been observed, where the top decays leptonically. All jets satisfy eq.~(\ref{eq:detector2}).}
\label{tab:bfrac}
\end{table}
Above, we have compared whether additional $b$ jets are harder than radiated light jets. One must also consider (in light of the first question above) the likelihood of observing events with radiated light jets or more than one $b$ jet, subject to the detector constraints. 

In panels (a) of Table~\ref{tab:bfrac} we show the probability to see a light-flavor jet in addition to the bottom jet from the top decay, for a number of different cases of the detector cuts
\begin{equation}
|\eta|<\eta_\text{cut} \qquad \qquad p_T>p_{T,\text{cut}},
\label{eq:detector2}
\end{equation}
and for leptonic decays of the top quark (i.e. where light jets can only come from QCD radiation). As expected, such a light QCD jet from soft and collinear jet radiation appears in
a sizeable fraction of the events. The lower percentages compared to
heavy sgluon or squark/gluino production~\cite{Plehn:2005cq,Plehn:2008ae,Alwall:2008qv} are due to
the fact that here we only consider central jets out to $|\eta|=2.5$.
This percentage increases with the hard scale in the process, such
that for a heavy Higgs more than half of all events show a central jet
within $|\eta|<2.5$ and above 25~GeV. Even above the typical range of
jets from $W$ decays, i.e. requiring $p_T > 45$~GeV, the fraction of events
with at least one QCD jet varies between 32\% and 42\%, depending on
the Higgs mass.

In panel (b) of Table~\ref{tab:bfrac} we show similar results, but for hadronic top decays, where light jets may arise either from QCD radiation or from the decay of the $W$ boson from the top quark. This increases the average
number of light-flavor jets per event by (roughly) two. Indeed, in panel (b) of
Table~\ref{tab:bfrac} we indeed see that the percentage of events with
jets above 25~GeV reaches close to 100\%. 

In panel (c) of Table~\ref{tab:bfrac} we show the probability of observing an additional $b$ jet, given that a hard $b$ jet has already observed. Again, as expected from collinear
radiation the percentage of events with two bottom jets decreases
sharply with more central or harder bottom jets. Only in the most
central phase space region with $p_T \sim 25$~GeV does the probability of
observing two bottom jets reach $\mathcal{O}(20\%)$. The
slightly higher probabilities for larger Higgs masses is an effect of
the generally hardened $p_T$ spectrum of the jet radiation. Considering the full detector region of eq.~(\ref{eq:detector}), 
the probability of seeing a second $b$ jet given a first is $17.8\%$ ($21.4\%$) for a light (heavy) charged Higgs boson. Again, from a QCD perspective this overall
percentage is very low~\cite{Plehn:2005cq,Plehn:2008ae,Alwall:2008qv} because we require
$|\eta|<2.5$ to allow for $b$ tagging, instead of the full jet range
$|\eta|<4.5$. However, this prediction is overly optimistic in the sense that we have assumed 100\% $b$-tagging efficiency. The numbers in Table~\ref{tab:bfrac} still have to be
multiplied with a $b$ tagging probability around $50\% - 60\%$ (or indeed even lower for the smaller $p_{t,cut}$ values), which
means that CMS and ATLAS can expect of the order of $10\%$ of their $H^-t$ events to include a second tagged bottom jet, if a first has been seen. This number is subject to higher order corrections to the $gg\rightarrow\bar{b}tH^-$ process~\cite{Dittmaier:2009np,Beccaria:2009my}, but these do not change things significantly.

The upshot of the above discussion is that one can observe two $b$ jets in only a small fraction of events. Furthermore, comparison of the transverse momentum distributions seems to indicate that additional $b$ jets are not substantially harder than radiated light jets. One may make this latter point more specific as follows. Consider that a hard $b$ jet has been observed. We may then ask the question: what is the probability that the hardest jet is a $b$ jet? This is shown in panels (a) and (b) of Table~\ref{tab:hardest} for the cases where the top decays leptonically and hadronically respectively. For leptonic decays, the probability is reasonably high (i.e. upwards of 80\%), reflecting the fact that for leptonic decays there is a hard $b$ jet from the top decay, but light jets arise only from QCD radiation. However, it is interesting (and perhaps surprising) to note that for $p_{T,cut}=25$ GeV and $\eta_{\text{cut}}=2.5$, a radiated light jet is harder than the primary $b$ jet in a sizeable fraction of cases (i.e. around 20\%). For hadronic decays (panel (b)), the probability that the hardest jet is a $b$ jet is much lower, given that there is competition from the light jets from the top decay. The numbers change only slightly in going to a higher Higgs boson mass.
\begin{table}
\begin{small} \begin{center}
\begin{tabular}{|cc||c|c|c|c|c||c|c|c|c|c|}
\hline
&&\multicolumn{5}{c||}{$\mh = 300$~GeV}
 &\multicolumn{5}{c|}{$\mh = 800$~GeV}\\
&&\multicolumn{5}{c||}{$\eta_\text{cut}$}
 &\multicolumn{5}{c|}{$\eta_\text{cut}$}\\

& $p_{T,\text{cut}}$
& 2.5 & 2.0 & 1.5 & 1.0 & 0.5
& 2.5 & 2.0 & 1.5 & 1.0 & 0.5\\
\hline
\hline
\multirow{4}{*}{(a)}
& 25~\gev & 80.6 &83.5  &86.7  &90.5  &95.0 
     & 77.4  &80.2   &83.8   &88.0   &93.2  \\
& 45~\gev & 85.2 &87.6  &90.2  &93.1  &96.3  
     &81.6  & 84.1  &87.1   &90.5   &94.5  \\
& 65~\gev & 89.3 &91.0  &93.1  &95.2  &97.4  
     &85.6  &87.7   & 90.2  &92.9  &95.7  \\
& 85~\gev &91.9  &93.4  &95.0  &96.5  &98.0  
     & 88.7  &90.4  &92.4  &94.6   &96.7  \\
\hline
\hline
\multirow{4}{*}{(b)}
& 25~\gev &46.3  &49.3  &53.9  &61.6  &75.8 
     & 46.8 & 49.5 &54.1  &61.4  &74.9 \\
& 45~\gev &55.7  &58.3  &62.3  &68.7  &80.1 
     & 54.0 &56.6  &60.6  &66.9  &78.2  \\
& 65~\gev & 68.8 &70.9  &73.9  &78.3  &85.7  
     &64.1  &66.3  &69.4  &74.3  &82.7  \\
& 85~\gev & 77.6 &79.2  &81.5  &84.4  &89.5  
     &72.0  &73.7  &76.2  &80.0  &86.3  \\
\hline
\hline
\multirow{4}{*}{(c)}
& 25~\gev &35.7  &34.5  &31.5  &26.7  &21.9 
     &32.7   &31.4   &28.8  &24.6  &19.8  \\
& 45~\gev &39.6  &39.0  &36.7  &31.6  & 25.3 
     & 36.3  &35.5  &33.1   &28.6   &22.4  \\
& 65~\gev & 43.6 & 43.9 & 42.3 & 37.5 &31.8  
     & 39.0  & 38.8  & 36.9  & 32.5  & 24.6 \\
& 85~\gev & 46.4 &47.9  &47.4  &44.0  &37.0  
     & 41.2 & 41.8  &40.6  &36.7  &27.9  \\
\hline
\hline
\multirow{4}{*}{(d)}
& 25~\gev &12.5  &11.9  &10.6  &8.7  &6.9
     &13.4  &12.8  &11.2  &9.3  &6.9 \\
& 45~\gev & 13.7 &12.9  &11.5  &9.0  &7.1 
     &14.5  &13.8  &12.1  &9.9  &7.2  \\
& 65~\gev & 17.1 &16.2  &14.2  &10.9  &7.8  
     & 16.8 &16.0  &13.9  &10.9  &7.5  \\
& 85~\gev & 20.1 &19.3  &17.3  &13.3  &8.9  
     &18.4  &17.6  &15.3  &11.9  &7.8  \\
\hline
\end{tabular}
\end{center} \end{small}
\caption{Probability (\%) that (a) the hardest jet is a $b$ jet, given that at least one $b$ jet has been observed, where the top decays leptonically; (b) the hardest jet is a $b$ jet, given that at least one $b$ jet has been observed, where the top decays hadronically; (c) probability that the two hardest jets are both $b$ jets, where the top decays leptonically; (d) probability that the two hardest jets are both $b$ jets, where the top decays hadronically.}
\label{tab:hardest}
\end{table}

Given that, as remarked above, one really wants to isolate two $b$ jets in order to reduce multijet backgrounds, the more experimentally relevant question is: given that one hard $b$ jet has been observed, how likely is it that
we find the second $b$ jet by asking for the two hardest jets in the
event? This probability is shown in panels (c) and (d) of Table~\ref{tab:hardest}, and is considerably below 50\% even for the best-suited phase space
regions. Thus, for central jets in the case of leptonic top decays we see
that the radiated light-flavor jets are typically harder than the
additional $b$ jet. This is simply a combinatorial effect --- from
collinear radiation we expect either one or three bottom jets, where
three bottom jets are strongly suppressed. For regular QCD radiation
we have many more diagrams leading to additional jets, so when we ask
for the hardest radiated jet it is usually one bottom jet vs. the
hardest of several light-flavor jets. As expected, this fraction drops
sharply for the (already experimentally less promising) hadronic top
decay signature, as can be seen in panel (d) of the table.

The above results demonstrate that isolating events in which there are two hard $b$ jets is difficult. There are not many events in which two or more observable $b$ jets are present, and their transverse momentum properties do not significantly distinguish additional $b$ jets from radiated light jets, even in the case of leptonic decays (when there are no hard light jets from the top quark decay). The above results, however, do not constitute a thorough phenomenological analysis, such that further investigation may be useful.

The preceding discussion is restricted to the region of large Higgs mass, where one does not have to worry about interference effects. One must also consider the regime of small Higgs mass, which is the subject of the next section. We will focus on interference issues, rather than properties of $b$ and light jets. However, it is worth bearing in mind that the above discussion would be significantly modified for small Higgs mass, where the second $b$ jet may indeed prove to be useful in reducing top pair production backgrounds (analogously to the $Wt$ process studied in~\cite{White:2009yt}).
\subsection{Small charged Higgs mass: $m_{H^-}<m_t$}
\label{resultssmall}
As described in section~\ref{interference}, when the charged Higgs boson mass is lower than the top mass, the $H^-t$ process at NLO interferes with top pair production at LO, with decay of the $\bar{t}$ particle into a charged Higgs boson and (anti)-bottom quark. It is still possible to construct an MC@NLO for $H^{-}t$ production, using the diagram removal and diagram subtraction definitions, first applied in the context of $Wt$ production in~\cite{Frixione:2008yi}. MC@NLO then gives a sensible result only subject to the approximation that the single and top pair processes can be added incoherently i.e. that the interference between the two processes can be neglected. This will not be true in general, but is likely to be true for analysis cuts used to increase the ratio of the single and top pair cross-sections. The DR and DS MC@NLO codes are defined independently of any subsequent analysis cuts, and the difference between these codes can be used to estimate the systematic uncertainty due to interference effects.

The object of this section is to both clarify and strengthen the above remarks, by comparing the DR and DS results for various charged Higgs boson masses. In order to verify that the DR and DS codes function as required, it is useful to have a continuous cut parameter that smoothly interpolates between the regions where the interference is expected to be small and large respectively. Following~\cite{Frixione:2008yi}, we introduce a transverse momentum veto on the second hardest $B$ hadron occurring in an event. That is, one first finds all $B$ hadrons satisfying the detector constraint
\begin{equation}
|\eta|<2.5,
\label{detcon}
\end{equation}
where $\eta$ is the pseudo-rapidity. Then events are rejected if a second hardest $B$ hadron is present, with transverse momentum satisfying:
\begin{equation}
p_t^b>p_{t,\text{veto}}.
\label{ptvetodef}
\end{equation}
The reasoning is that top pair-like events at the LO parton level contain two $b$ quarks in the final state, whereas $H^-t$-like events contain only one. At NLO plus parton shower level, one still expects top pair-like events to contain roughly two observable $B$ hadrons, and single top-like events to have less than two. Thus, for small values of the $p_{t,\text{veto}}$, the top pair process is efficiently reduced relative to the single top mode. Large values of the veto ($p_{t,\text{veto}}\rightarrow\infty$) correspond to no constraint, and an interference that is not expected to be small in general. 

Note that there are different choices in how to apply this veto. One can apply it even in the pure NLO calculation, by using $b$ quarks rather than $B$ hadrons, although there are difficulties interpreting this phenomenologically (see the detailed discussion in~\cite{Frixione:2008yi}). In the parton shower approach, one could also apply the veto at the level of $b$ jets for example. However, the aim here is merely to check the mutual validity of the DR and DS procedures, rather than to be phenomenologically complete. Also, applying the veto in the same way as in the $Wt$ case of~\cite{Frixione:2008yi} allows one to compare the size of interference effects in both cases.

For renormalization / factorization scale choices and top quark mass fixed as above, the level of disagreement between DR and DS will depend in general on the charged Higgs boson mass as well as the transverse momentum veto. Thus, we evaluate cross-sections for a number of choices of $m_{H^-}<m_t$. Results are shown in table~\ref{DRvDSresults}, also for a number of choices of $p_{t,\text{veto}}$. The quoted uncertainties are due to variation of the common renormalization and factorization scale by a factor of two. This uncertainty can be larger if the scales are allowed to vary independently. 
\begin{table}
\begin{center}
\begin{tabular}{|cc||c|c|c|c|c|}
\hline
\multicolumn{2}{|c||}{$m_{H^-}$} & \multicolumn{5}{c|}{$p_{t,\text{veto}}$}\\
& & 10 & 30 & 50 & 70 & $\infty$\\
\hline
\hline
\multirow{2}{*}{100}& DR &$1.607^{+0.043}_{-0.036}$ &$1.949^{+0.058}_{-0.039}$ &$2.069^{+0.057}_{-0.039}$ &$2.108^{+0.055}_{-0.041}$ &$2.129^{+0.054}_{-0.044}$ \\
& DS & $1.544^{+0.053}_{-0.048}$ & $1.812^{+0.080}_{-0.058}$ & $1.857^{+0.100}_{-0.088}$ & $1.870^{+0.103}_{-0.103}$  & $1.870^{+0.108}_{-0.113}$\\
& Ratio & 1.041 & 1.076 & 1.114 & 1.127 & 1.139\\
\hline
\multirow{2}{*}{120}& DR &$1.311^{+0.033}_{-0.026}$ & $1.615^{+0.033}_{-0.041}$ & $1.702^{+0.035}_{-0.044}$ & $1.725^{+0.035}_{-0.045}$ & $1.735^{+0.036}_{-0.046}$\\
& DS & $1.252^{+0.046}_{-0.043}$ & $1.478^{+0.063}_{-0.085}$ & $1.518^{+0.076}_{-0.099}$ & $1.528^{+0.079}_{-0.102}$ & $1.529^{+0.083}_{-0.105}$\\
& Ratio & 1.047 & 1.093 & 1.121 & 1.129 & 1.135\\
\hline
\multirow{2}{*}{140}& DR & $1.087^{+0.023}_{-0.021}$ & $1.344^{+0.034}_{-0.031}$ & $1.389^{+0.039}_{-0.032}$ & $1.398^{+0.039}_{-0.032}$ & $1.401^{+0.040}_{-0.033}$\\
& DS & $1.037^{+0.035}_{-0.039}$ & $1.227^{+0.055}_{-0.072}$ & $1.265^{+0.059}_{-0.077}$ & $1.273^{+0.060}_{-0.079}$ & $1.274^{+0.062}_{-0.080}$\\
& Ratio & 1.048 & 1.095 & 1.098 & 1.098 & 1.100\\
\hline
\multirow{2}{*}{160}& DR & $0.939^{+0.025}_{-0.031}$ & $1.078^{+0.032}_{-0.038}$ & $1.090^{+0.034}_{-0.038}$ & $1.093^{+0.034}_{-0.038}$ & $1.094^{+0.035}_{-0.038}$\\
& DS & $0.897^{+0.034}_{-0.040}$ & $1.031^{+0.044}_{-0.045}$ & $1.045^{+0.045}_{-0.045}$ & $1.048^{+0.046}_{-0.045}$ & $1.049^{+0.046}_{-0.045}$\\
& Ratio & 1.047 & 1.046 & 1.043 & 1.043 & 1.043\\
\hline
\end{tabular}
\caption{Total MC@NLO cross-section results (in pb) for $H^-t$ cross-sections subject to the transverse momentum veto of eq.~(\ref{ptvetodef}), with parameters as described in the text. Both the DR and DS results are shown, and errors shown are due to variation of the common renormalization and factorization scale by a factor of two. The statistical error on the ratio values is less than a percent.}
\label{DRvDSresults}
\end{center}
\end{table}
One sees that the level of agreement between DR and DS is around 4-5\% at $p_{t,\text{veto}}=10$ GeV. This is higher than in the case of $Wt$ production, indicating that interference effects are larger in the present case (they also depend upon the Higgs mass of course). Nevertheless, the DR and DS results overlap within the scale variation uncertainty. The results with no veto applied show a marked disagreement (greater than 10\% in some cases). This shows that interference with $t\bar{t}$ production is indeed a problem in some phase space regions, as is to be expected. Interestingly, for the value of the charged Higgs mass that lies closest to the top quark mass, the results agree well even when no veto is applied. We will discuss the threshold region in more detail in the following section.

\begin{figure}
\includegraphics[width=0.48\textwidth]{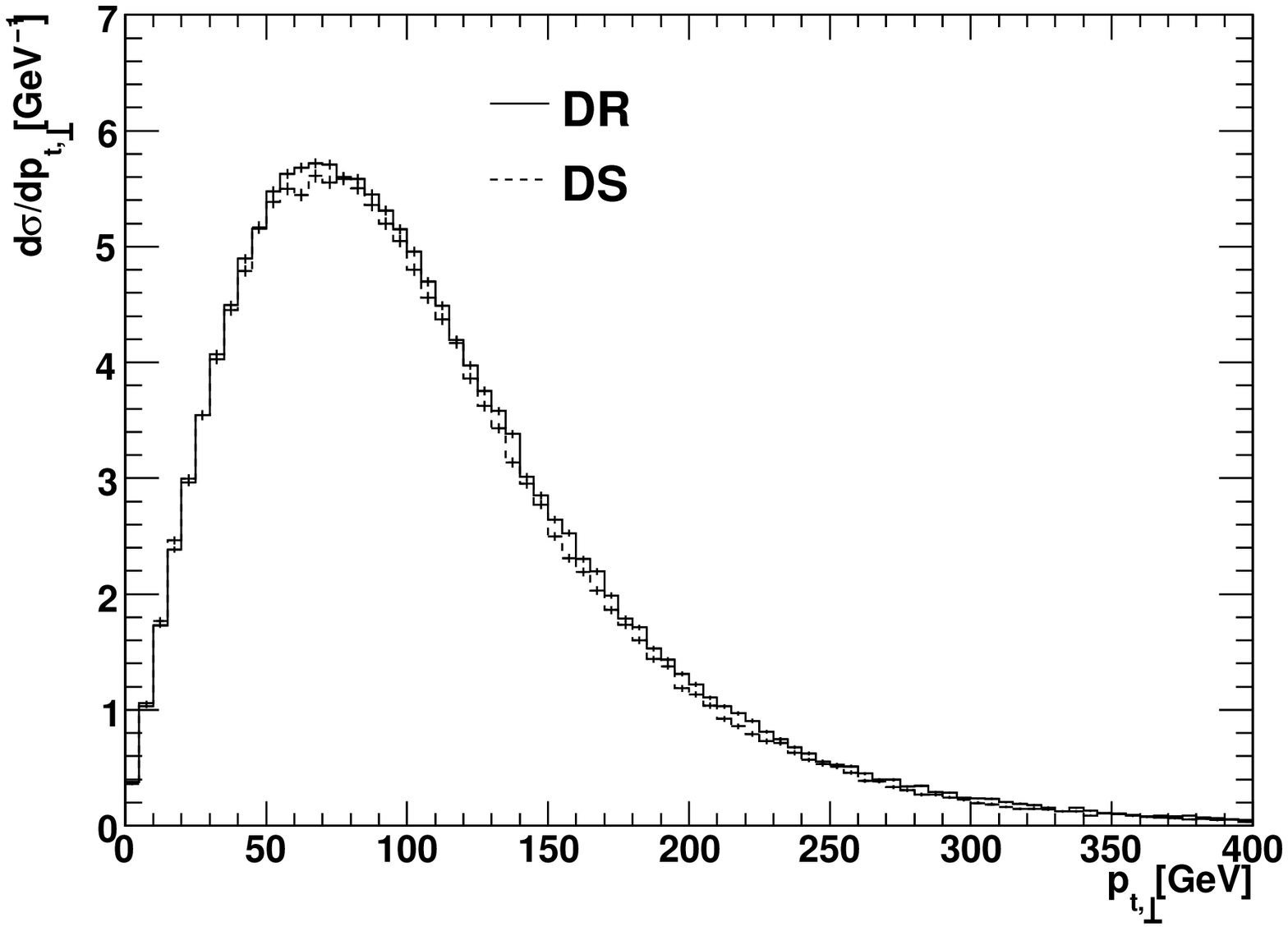}
\includegraphics[width=0.48\textwidth]{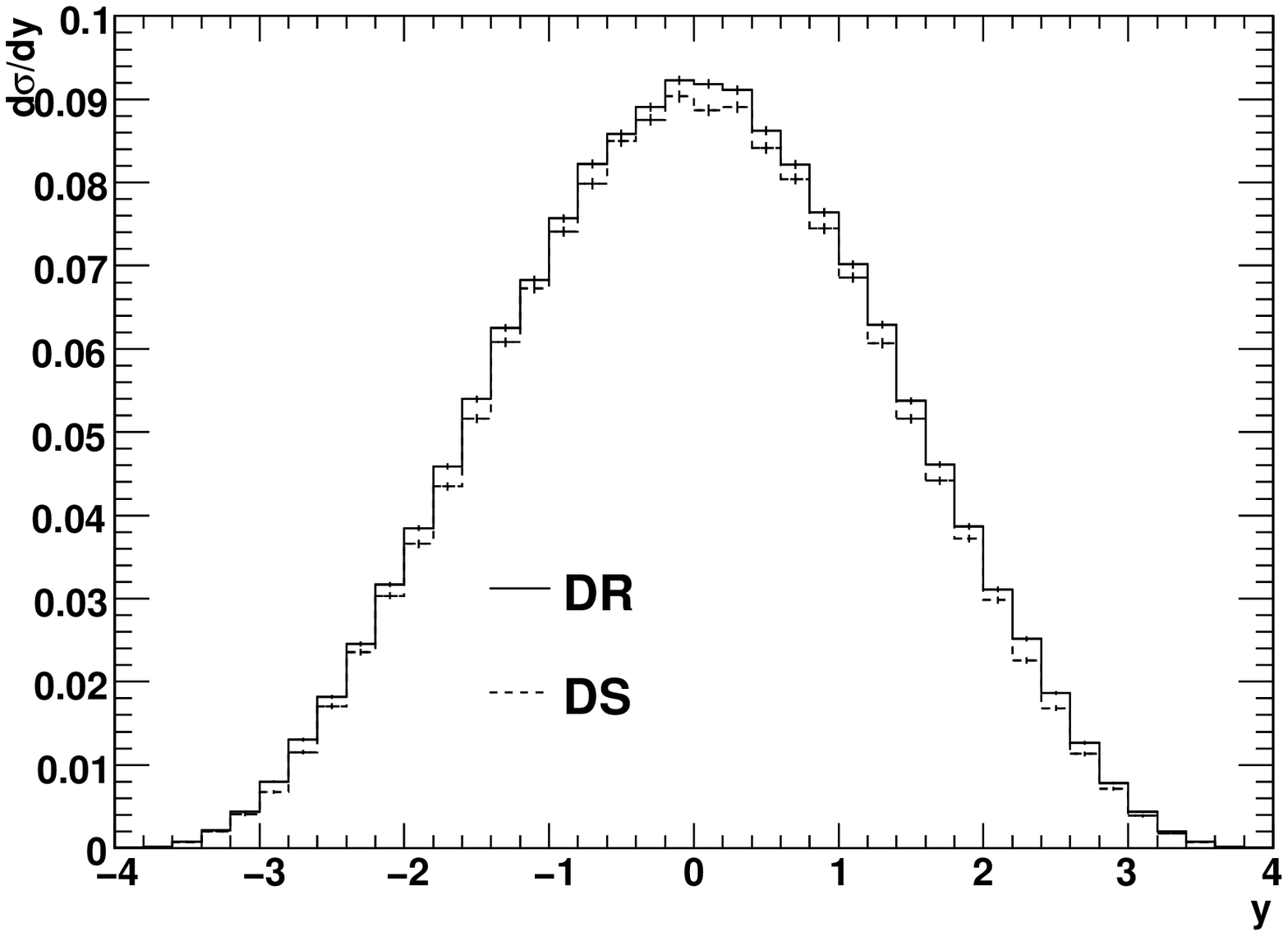}
\includegraphics[width=0.48\textwidth]{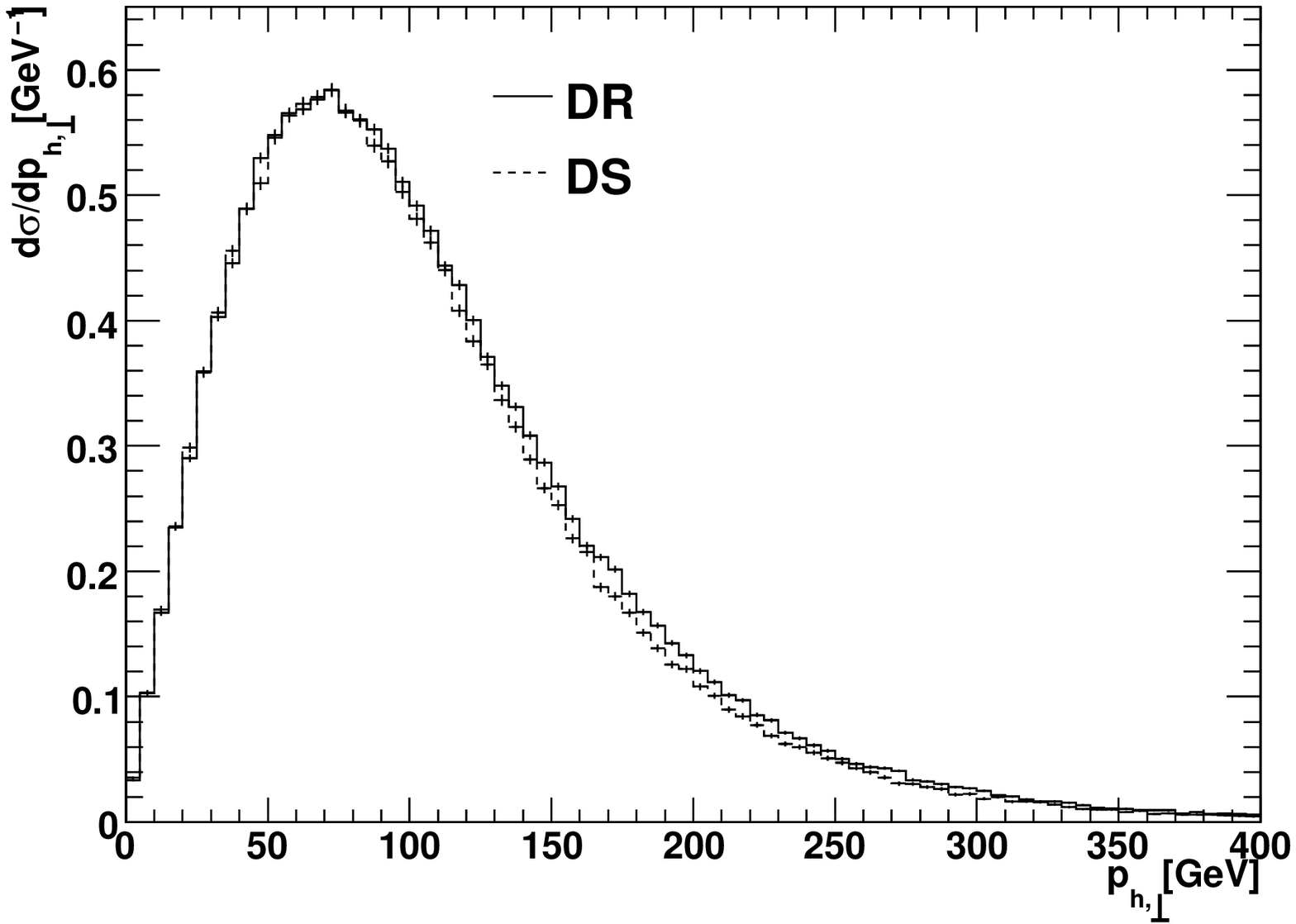}
\includegraphics[width=0.48\textwidth]{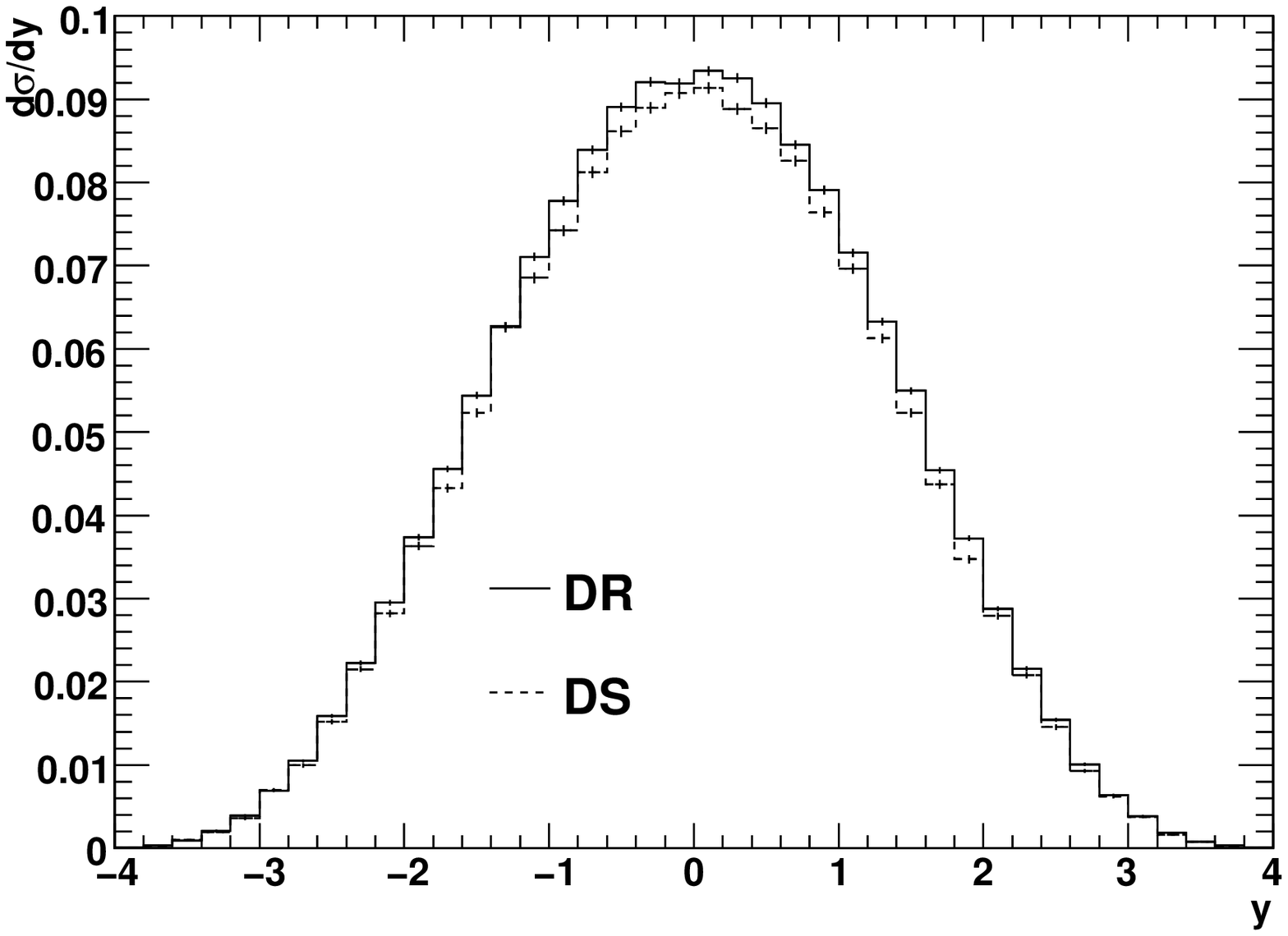}
\includegraphics[width=0.48\textwidth]{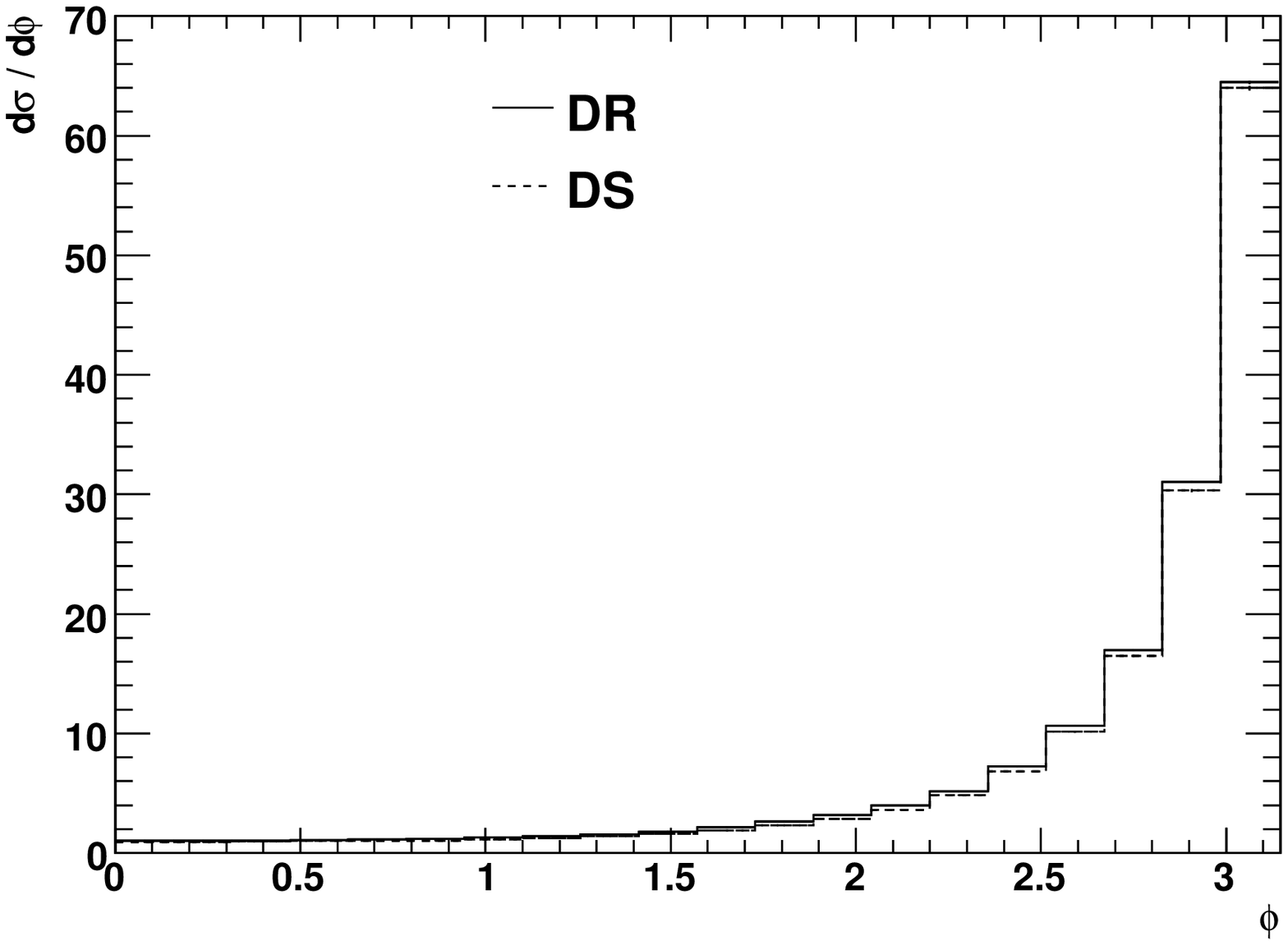}
\caption{Comparison of DR and DS results, for $m_{H^-}=100$ GeV, and other parameters as outlined in the text. Shown are the transverse momentum and rapidity distributions of the top quark (upper line) and Higgs boson (second line), as well as the azimuthal angle between the top quark and charged Higgs boson (lower line). Error bars denote statistical uncertainties.}
\label{DRvDSfigs}
\end{figure}

Total cross-sections are only part of the story. One must also check that interference effects are small locally in phase space, and this is possible by comparing kinematic distributions obtained from both the DR and DS codes. Examples are shown in figure~\ref{DRvDSfigs} for $p_{t,\text{veto}}=10$ GeV, and $m_{H^-}=100$ GeV (results are representative of other charged Higgs masses). One indeed sees good agreement in the observables considered.
Some comments are in order. Firstly, we have here shown only a few observables, with a particular cut designed to reduce the interference with top pair production (i.e. the transverse momentum veto). A more realistic analysis designed to isolate the charged Higgs production process would use different cuts to reduce the interference, such as jet vetoes (see e.g. the $Wt$ analysis of~\cite{White:2009yt}). It may still be possible to define observables in such cases that are sensitive to interference effects, and whether or not this is the case can be ascertained by comparing the output of the DS and DR predictions. Nevertheless, the results here indicate that the incoherence of single and top pair-instigated charged Higgs production is a reasonable approximation, over a sufficiently large region of the phase space for MC@NLO to provide an accurate description when possible (as was found to be the case in $Wt$ production in~\cite{White:2009yt}). In the next section, we explain how to interpret the behavior of the MC@NLO calculation across the threshold $m_{H-}= m_t$.
\subsection{Threshold behavior}
\label{threshold}
In the previous sections, we have outlined the calculation of the $H^-t$ production process up to NLO, and described the interface to a parton shower with the MC@NLO formalism. For $m_{H^-}>m_t$ this is straightforward, whereas for $m_{H-}<m_t$ a problem arises due to interference with top pair production. The aim of this section is to clarify any possible confusion in the behavior of the cross-section across the top mass threshold.

A consequence of the subtraction inherent in the DR and DS definitions is that the MC@NLO prediction for $H^-t$ itself is not continuous across the threshold $m_{H^-}=m_t$. Nor should it be, given that the sum of the single and top pair processes is continuous, and the top pair contribution is subtracted out in either DR or DS. The result is that there is a discontinuity in the $H^-t$ calculation (defined using either DR or DS) as $m_{H-}$ tends towards $m_t$ from below. This discontinuity would (in MC@NLO) be filled in by the addition of top pair contributions, which in any given analysis must be added as a background. 

\begin{figure}
\begin{center}
\scalebox{0.8}{\includegraphics{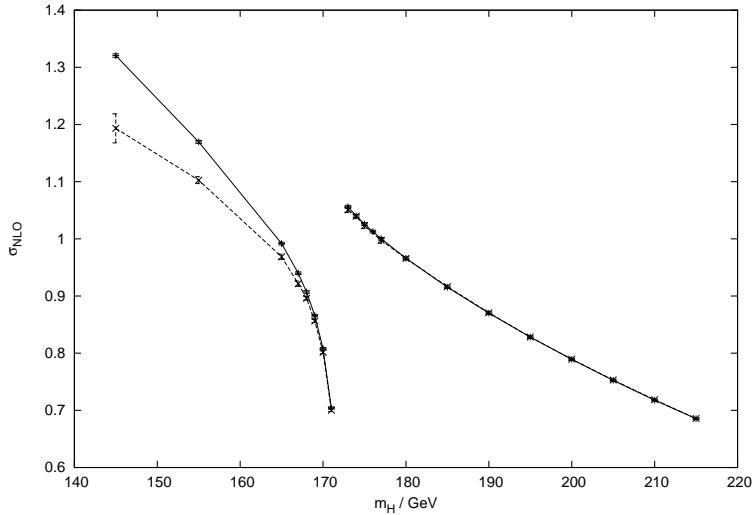}}
\caption{The total NLO cross-section for $H^-t$ production, shown for DR (solid) and DS (dashed), for a top mass of 172 GeV. The error bars correspond to statistical uncertainties. The discontinuity is due to subtraction of resonant contributions for $m_{H^-}<m_t$, and would be filled in after adding in the top pair production background.}
\label{threshfig}
\end{center}
\end{figure}

This is depicted explicitly in figure~\ref{threshfig}, which shows the fully inclusive total $H^{-}t$ cross-section calculated using DR and DS as a function of the charged Higgs boson mass, with all other parameters set as above (in particular $m_t=172$ GeV). The curves disagree for the low mass region, as expected if no analysis cuts are applied to reduce the interference with top pair production. The apparently unphysical behavior near $m_{H^-}=m_t$ is purely a consequence of the fact that MC@NLO is calculating a qualitatively different cross-section on either side of this value. For the high Higgs mass region $m_{H^-}>m_t$, the curves agree due to the fact that no subtraction is applied.

We stress that no interpolation is necessary, or indeed meaningful, between the calculations on either side of the mass threshold. Furthermore, the discontinuity is not problematic in any real analysis, in which the $H^-t$ prediction must be combined with the corresponding top pair process (a discussion of threshold behavior can also be found in the context of a strictly NLO calculation in~\cite{Berger:2003sm}, where however the DR and DS definitions have not been used). 
\section{Conclusion}
\label{conclusion}
In this paper, we have considered the process of charged Higgs boson production in association with a top quark. This is an important scattering process at the LHC, given that charged Higgs bosons generically occur in extensions to the Standard Model. In order to increase the accuracy with which this process can be calculated, we have implemented it in the MC@NLO framework for combining fixed order matrix elements at next-to-leading order with a parton shower. 

The details of the MC@NLO calculation are different in the two kinematic regions in which the charged Higgs boson mass is less than and greater than the top mass respectively. For $m_{H^-}>m_t$, the NLO calculation of the $H^-t$ process is well-defined, and the implementation in MC@NLO follows the procedure adopted in other single top (and indeed non-top related) processes~\cite{Frixione:2002ik,Frixione:2003ei,Frixione:2005vw}. 

In this high Higgs mass region, we presented example kinematic distributions involving the top and charged Higgs bosons, which demonstrated the expected differences between the fixed order and parton showered approaches. We also looked at the properties of $b$ jets in addition to the top decay, to see if these are in any way different to radiated light jets (motivated by previous analyses which aim to exploit such differences). The conclusion is that radiated $b$ jets are not substantially harder than light jets. Furthermore, the fraction of events containing two $b$ jets is small.

For the low Higgs mass region $m_{H^-}<m_t$, the NLO cross-section for $H^-t$ production becomes ill-defined due to interference with top pair production at LO, with decay of the $\bar{t}$ to a charged Higgs boson and $\bar{b}$ quark. There is then some ambiguity in how to proceed, and we proceed by analogy with the $Wt$ case discussed in~\cite{Frixione:2008yi,White:2009yt}. That is, one may approximate the sum of the single and top pair production processes by an incoherent sum, subject to adequate cuts on the phase space. We produced two MC@NLO implementations using the diagram subtraction (DS) and diagram removal (DR) approaches, such that the difference between the two codes provides an estimate of the systematic uncertainty due to interference effects. The resulting software can then be used in the low Higgs mass region to efficiently generate $H^-t$-like events.

Using the DR and DS codes we find that, whilst the interference is sizeable when no cuts are applied (as is to be expected), the DR and DS definitions agree well when a suitable cut is applied. Here we used the simple example of a transverse momentum veto on the second hardest $B$ hadron in an event. The results are similar to the $Wt$ case of~\cite{Frixione:2008yi} (albeit with a larger interference effect seen in the present case), and therefore suggest that isolation of the charged Higgs signal is possible in more realistic analyses (based on e.g. $b$ jet vetoes). We postpone more detailed phenomenology to future work.

\section{Acknowledgments}
 MK and CW thank Karol Kovarik for his collaboration during
 the initial stages of this project.
Their work is supported
 by the grant ANR-06-JCJC-0038 and the Theory-LHC-France
 initiative of CNRS/IN2P3.
 This research has been supported by the Foundation for Fundamental Research of Matter (FOM) as part of the programme TPP (Theoretical Particle Physics in the era of the LHC, FP 104) and the National Organization for Scientific Research (NWO). CDW was supported by the Marie Curie grant ``TOP@LHC'' as well as by the STFC postdoctoral fellowship ``Collider Physics at the LHC''. He thanks Peter Richardson for help with HERWIG, and we are also grateful to Fabio Maltoni for discussions relating to the renormalization of the Yukawa coupling.  

\appendix
\section{Renormalization of Yukawa couplings}
\label{renormy}
In this appendix we derive the counterterms resulting from the renormalization of the charged Higgs Yukawa couplings. We first examine the case of a scalar neutral Higgs boson, before considering the example of a type II two-Higgs-doublet model. Similar remarks would apply in the case of a type I model. Our presentation follows that of the neutral Higgs boson case presented in~\cite{Campbell:2002zm}.

The Yukawa coupling of the Higgs boson to quarks in the Standard Model is given by:
\begin{equation}
{\cal L}_{\text{Yuk.}}=-\frac{1}{\sqrt{2}}\lambda (v+H) \bar{Q}_L Q_R,
\label{yuk0}
\end{equation}
where $v$ is the vacuum expectation value of the Higgs boson field $H$, and $Q_{L,R}$ are the left and right-handed quark fields. From the above Lagrangian one sees that the bare quark mass and Yukawa couplings are related by:
\begin{equation}
m_0=\frac{\lambda_0 v}{\sqrt{2}}.
\label{mf}
\end{equation}
At one-loop order the bare mass and Yukawa couplings are renormalized according to:
\begin{align}
\lambda_0&=\mu^\epsilon \lambda(1+\delta \lambda)\label{y0renorm};\\
m_0&=m+\delta m\label{mrenorm2},
\end{align}
where $\delta\lambda$ and $\delta m$ are the appropriate counterterms, and $\lambda$ and $m$ (scheme-dependent) renormalized quantities. One may choose a renormalization scheme such that the Higgs VEV $v$ is not renormalized at one loop order. Then one finds
\begin{equation}
\delta \lambda=\frac{\delta m}{m},
\end{equation}
i.e. that the Yukawa coupling renormalization is simply related to the mass renormalization. Note that $\delta m\propto m$, such that the Yukawa coupling counterterm is independent of the quark mass. In the $\overline{\text{MS}}$ scheme, this has the form
\begin{equation}
\delta\lambda^{\overline{\text{MS}}}=\frac{C_F\alpha_S}{4\pi}\left(-\frac{3}{\epsilon}+3\gamma_E-3\ln4\pi\right).
\label{lammsbar}
\end{equation}

We now consider the case when two Higgs doublets are present, and focus on the case of a type II two-Higgs-doublet model. Focussing on one quark generation, the coupling of the Higgs doublets to the fermions is given by (see e.g.~\cite{Djouadi:2005gj})
\begin{equation}
{\cal L}_{Yuk.}^{\text{MSSM}}=-\epsilon_{ij}\left[(\lambda_b + \delta\lambda_b )\bar{b}_R H_1^i Q_L^j+(\lambda_t+\delta\lambda_t)\bar{t}_RQ_L^jH_2^i+\ldots\right],
\label{lyuk2}
\end{equation}
where the ellipsis denotes terms obtained from those shown by Hermitian conjugation. Here $R$ and $L$ denote right- and left-handed quarks respectively, and $i$ and $j$ are flavor indices. The Higgs and fermion SU(2) doublets are given by
\begin{equation}
H_1=\left(\begin{array}{c}H_1^0\\H_1^-\end{array}\right),\quad 
H_2=\left(\begin{array}{c}H_2^+\\H_2^0\end{array}\right),\quad
Q_L=\left(\begin{array}{c}t_L\\b_L\end{array}\right),
\label{doublets}
\end{equation}
where $H_a^0$ and $H_a^\pm$ are the neutral and charged Higgs boson respectively. Expanding each Higgs doublet about its vacuum expectation value corresponds to the replacements (including conventional factors of $\sqrt{2}$)
\begin{align}
H_1\rightarrow\frac{1}{\sqrt{2}}\left(\begin{array}{c}v_1+H_1^0+iP_1^0\\H_1^-\end{array}\right),\quad H_2\rightarrow\frac{1}{\sqrt{2}}\left(\begin{array}{c}H_2^+\\v_2+H_2^0+iP_2^0 \end{array}\right),
\label{VEVs}
\end{align}
where $P_a^0$ correspond to the pseudo-scalar Higgs particles and Goldstone bosons. Substituting eq.~(\ref{VEVs}) into eq.~(\ref{lyuk2}), one may then apply an analogous argument to the Standard Model case to each Higgs doublet separately, to obtain the relations
\begin{equation}
\delta \lambda_b=\frac{\delta m_b}{m_b},\quad \delta\lambda_t=\frac{\delta m_t}{m_t}.
\label{yukrenorm2}
\end{equation}
It is conventional to define the parameters $v$ and $\beta$ parameter from the two Higgs VEVs i.e.
\begin{equation}
\tan{\beta}=\frac{v_2}{v_1}=\frac{v\sin\beta}{v\cos\beta}.
\label{tanbeta}
\end{equation}
Then the coupling of the physical charged Higgs field to fermions can be written
\begin{equation}
G_{H^- t\bar{b}}=-\frac{i}{\sqrt 2}{v}V_{tb}\Big[m_b\tan\beta(1-\gamma_5)+m_t\cot\beta(1+\gamma_5)\Big],
\label{lyuk3}
\end{equation}
where the appropriate CKM matrix element has been factored out. This has the form
\begin{equation}
G_{H^- t\bar{b}}=iV_{tb}(a-b\gamma_5),
\label{G2}
\end{equation}
where 
\begin{equation}
a=\frac{1}{\sqrt{2}}\left(m_b\tan\beta+m_t\cot\beta\right),\quad b=\frac{1}{\sqrt{2}}\left(m_b\tan\beta-m_t\cot\beta\right).
\label{aandb}
\end{equation}
From the above remarks, these clearly have the corresponding one-loop renormalization
\begin{equation}
a_0=a(1+\delta a), \quad b_0=b(1+\delta b),\quad \delta a=\delta b=\frac{\delta m}{m},
\label{aandbcounter}
\end{equation}
where $m$ is the mass of either the top or down-type quark, given the independence of the final result on the mass.

\section{Calculation of $\tilde{{\cal M}}$}
\label{app-hel}
 The counterterms to real-emission matrix elements in subtraction formalisms
 have to cancel {\em locally} the relevant divergences, in order for the 
 numerical integration to be well-defined and stable.
 The case of the collinear branching of a gluon (in the initial or final
 state, with the gluon being spacelike and timelike respectively) merits
 some attention because it has a non-trivial local azimuthal dependence,
 that vanishes when the azimuthal integration is carried out.
 As shown in ref.~\cite{Frixione:1995ms}, in the FKS subtraction formalism the azimuthal-dependent 
 part of the local collinear counterterm is quite simple, and is formally 
 identical to the better-known azimuthal-independent one: both have the 
 structure of a kernel (that depends only on the parton identities, but 
 not on the hard process), times a short-distance, Born-like cross section.

To be more specific, consider the process
\begin{equation}
\alpha(p_1)+\beta(p_2)\longrightarrow X(K) +\delta(k),
\label{proc}
\end{equation}
where Greek letters represent (massless) partons, and $X$ any other particles that may be present (i.e. in the present case,
$X\equiv t+H^-$). In the collinear limit $p_2\parallel k$, the squared matrix element for the above process is given by
(see eq. (B.41) of~\cite{Frixione:1995ms})
\beq
\mat(p_1,p_2)\,\stackrel{p_2\parallel k}{\longrightarrow}\,
\frac{4\pi\as}{k\mydot p_2}\left[P(z)\mat^{(0)}(p_1,zp_2)+
Q(z)\tilde{\mat}(p_1,zp_2)\right]\,,
\label{Mclimit}
\eeq
where $P$ are the usual Altarelli-Parisi kernels, $Q$ are other universal
kernels (given at the leading order in eqs.~(B.42)--(B.45) of 
ref.~\cite{Frixione:1995ms} for initial-state collinear splittings, 
and in eqs.~(B.31)--(B.34) of that paper for final-state collinear splittings; 
these kernels are thus different for spacelike and timelike branchings already 
at the leading order), $\mat^{(0)}$ is the relevant Born contribution,
and $\tilde{\mat}$ is a Born-like function, which however keeps track
of the azimuthal correlations in the branching process. The contribution
of $Q\tilde{\mat}$ vanishes upon integration over the azimuthal angle
of the branching,
which is why this term can be neglected in the analytical computation of 
the collinear divergences in $4-2\epsilon$ dimensions. Locally, it is 
different from zero if the parton involved in the branching which enters
the hard reaction is a gluon, and therefore
needs to be taken into account for the construction of the local counterterms
in a (efficient) numerical NLO program.

The helicity interference amplitude is given by
\begin{equation}
\tilde{\cal M}={\cal F}\real\left\{\frac{\langle k p_2\rangle}{[k p_2]} 
\ap^{(0)\dag}\am^{(0)}\right\},
\label{mt1}
\end{equation}
and in the case of $Wt$ production in~\cite{Frixione:2008yi}, we calculated this in the helicity
formalism. Here we adopt a different (and somewhat quicker) approach, involving projection of the helicity dependent
Born amplitude with relevant tensors. 

Let ${\cal M}^{(0)}_{\mu\nu}$ be the Born amplitude of figure \ref{Bornfig} before contraction with the gluon 
polarization tensor $(-g^{\mu\nu})$. Then the helicity interference is given by~\cite{Mangano:1991jk}
\begin{equation}
\label{helterm}
\tilde{\cal M}=\left(-\frac{g^{\mu\nu}}{2}+\frac{p_1^\mu p_2^\nu+p_2^\mu p_1^\nu}{2p_1\cdot p_2}+\frac{k_\perp^\mu k_\perp^\nu}{k_\perp^2}\right){\cal M}^{(0)}_{\mu\nu},
\end{equation}
where $k_\perp$ is the transverse component of the gluon momentum entering eq.~(\ref{proc}). Carrying out the calculation in the present case gives
\begin{align}
\tilde{\cal M}&=\frac{N_cC_Fg_s^4\left(|a|^2+|b|^2\right)}{4N_c(N_c^2-1)}\notag\\
&\quad\times\frac{4 (2\cos^2\phi-1) (m_t^2 - m_{H^-}^2) (m_t^2 (m_{H^-}^2 - u) + 
   u (-m_{H^-}^2 + s + u))}{s (m_t^2 - u)^2},
\end{align}
where $\phi$ is the gluon azimuthal angle, and we have defined the Mandelstam invariants
\begin{equation}
s=(p_1+p_2)^2,\quad t_1=t-m_t^2=(k_1-p_1)^2-m_t^2,\quad u_1=u-m_t^2=(k_2-p_1)^2-m_t^2.
\label{invars}
\end{equation}
One sees immediately that this term vanishes upon integration over $\phi$.

\bibliographystyle{JHEP}
\bibliography{refs.bib}

\end{document}